\documentclass[aps,prl,10pt,twocolumn,superscriptaddress,longbibliography,nobalancelastpage]{revtex4-1}

\usepackage{amsmath}
\usepackage{amssymb}
\usepackage{graphicx}
\usepackage{dsfont}
\usepackage{braket}
\usepackage{bm}
\usepackage{wrapfig}
\usepackage{kantlipsum}
\usepackage{mathtools}
\usepackage[colorlinks=true,linkcolor=red,urlcolor=magenta,citecolor=blue]{hyperref}
\usepackage{tikz}
\usepackage[english]{babel}

\definecolor{g}{rgb}{.1,0.4,.1} 
\definecolor{b}{rgb}{0,0.2,1}
\definecolor{rouge}{rgb}{0.82,0.,0.}
\definecolor{vert}{rgb}{0.,0.82,0.}
\definecolor{orange}{rgb}{1,0.5,0.}
\definecolor{bleu}{rgb}{0.,0.,0.82}
\definecolor{m}{rgb}{0.82,0.,0.82}
\definecolor{vert2}{rgb}{0.,0.5,0.}
\definecolor{rougeclair}{rgb}{1.0,0.7,0.7}
\definecolor{gris}{rgb}{.8,.8,.8} 

\newcommand{\e}{\mathrm{e}}
\renewcommand{\d}{\mathrm{d}}
\newcommand{\one}{\mathds{1}}
\renewcommand{\dag}{^\dagger}

\newenvironment{diagram}
{
\begin{tikzpicture}[baseline = (X.base),every node/.style={scale=0.8},scale=.45]
}
{
\end{tikzpicture}
}

\begin{document}

\title{Quasiparticles in quantum spin chains with long-range interactions}

\author{Laurens Vanderstraeten}
\email{laurens.vanderstraeten@ugent.be}
\affiliation{Department of Physics and Astronomy, University of Ghent, Krijgslaan 281, 9000 Gent, Belgium}

\author{Maarten Van Damme}
\affiliation{Department of Physics and Astronomy, University of Ghent, Krijgslaan 281, 9000 Gent, Belgium}

\author{Hans Peter B\"{u}chler}
\affiliation{Institute for Theoretical Physics III and Center for Integrated Quantum Science and Technology, University of Stuttgart, 70550 Stuttgart, Germany}

\author{Frank Verstraete}
\affiliation{Department of Physics and Astronomy, University of Ghent, Krijgslaan 281, 9000 Gent, Belgium}
\affiliation{Vienna Center for Quantum Science and Technology, Faculty of Physics, University of Vienna, Boltzmanngasse 5, 1090 Vienna, Austria}

\begin{abstract}
We study quasiparticle excitations for quantum spin chains with long-range interactions using variational matrix product state techniques. It is confirmed that the local quasiparticle ansatz is able to capture those excitations very accurately, even when the correlation length becomes very large and in the case of topological nontrivial excitation such as spinons. It is demonstrated that the breaking of the Lieb-Robinson bound follows from the appearance of cusps in the dispersion relation, and evidence is given for a crossover between different quasiparticles as the long-range interactions are tuned.
\end{abstract}

\maketitle

In the last decades the research on low-dimensional quantum matter has exploded thanks to experimental advances in atomic, molecular and optical physics on the one hand, and a deepened theoretical understanding of the quantum many-body problem on the other. In a recent development the rich variety of exotic strongly-correlated quantum phases has increased even further as a result of the experimental manipulation of quantum spin systems with long-range interactions. These appear as, e.g., van der Waals interactions between Rydberg atoms \cite{Schauss2012}, dipole-dipole interactions between atoms and molecules \cite{Peter2012}, and can even be tuned in trapped-ion setups \cite{Richerme2014, Jurcevic2014}. On the theoretical side it has been realized that qualitively new phenomena can occur in the presence of long-range interactions. For example, the Mermin-Wagner theorem \cite{Mermin1966} for one-dimensional quantum systems can be avoided in the presence of long-range interactions, making the spontaneous breaking of continuous symmetries possible down to zero temperature \cite{Maghrebi2017}. An even more fundamental question concerns the fate of the Lieb-Robinson bound \cite{Lieb1972} in systems with long-range interactions. This bound implies that local quantum lattice systems exhibit a linear light cone out of which time-dependent correlation functions are suppressed exponentially, and is the starting point for proving e.g. area laws for entanglement entropy \cite{Hastings2007}. Although weaker versions of the Lieb-Robinson bound have been proven for power-law decaying interactions \cite{Hastings2006a, Foss-Feig2015}, it remains an open question whether a (non-linear) light cone emerges for generic long-range interacting models \cite{Hauke2013a,Maghrebi2016,Buyskikh2016}.
\par For local Hamiltonians the key to understand the low-energy dynamics of generic spin chains is traditionally provided by the notion of quasiparticles. Indeed, it was recently proven (using Lieb-Robinson bounds) that elementary excitations of gapped local Hamiltonians can be created out of the ground state by a momentum superposition of a local operator \cite{Haegeman2013}. This suggests that all low-energy excitations can be understood as localized perturbations or quasiparticles on top of a possibly strongly-correlated ground state. Dynamical properties such as structure factors, response functions, and quench dynamics can then be understood through the properties of these quasiparticles. In this light the emergence of a light cone is understood by a ballistic spreading of quasiparticles with a given characteristic velocity \cite{Calabrese2006, Calabrese2007a}.
\par When the interactions are no longer local, this quasiparticle picture is no longer guaranteed to capture the low-energy dynamics. Indeed, because of the absence of sharp Lieb-Robinson bounds, sub- or super-ballistic propagation through the system is possible, and it is a priori not clear that quasiparticles provide an understanding of these phenomena. Recently, though, there have been a few attempts in that direction. Perturbative continuous unitary transformations yield a quasiparticle description of excitations \cite{Fey2016}, but require a trivial point to perturb from in the same phase. In a semiclassical regime, spin-wave theory provides a good approximation for the low-energy dynamics in terms of the propagation of free magnons \cite{Cevolani2016, Cevolani2017, Frerot2018}.  Alternatively, quadratic field theories might provide an effective description of the system's dynamics in certain regimes \cite{Maghrebi2016}. In these approaches the non-linear behaviour of the light-cone can be traced back to non-analyticities in the quasiparticle dispersion relation. Still, a generic framework for strongly-interacting systems in the quantum regime is lacking.
\par For local interactions, such a framework \cite{Vanderstraeten2015a, Vanderstraeten2016} was developed using the language of matrix product states, and provides a comprehensive understanding of low-energy excitations in spin chains as quasiparticles. In this paper we extend this framework to the case of long-range interactions, and show that it continues to capture the relevant low-energy degrees of freedom.

\par\noindent\emph{The quasiparticle ansatz}--- %
Our framework starts from the formalism of matrix product states \cite{Verstraete2008a, Schollwock2011a, Orus2013}, a class of states that parametrizes the ground states of generic quantum spin chains. In the thermodynamic limit, a matrix product state can be represented graphically as
\begin{equation} 
\ket{\Psi(A)} =  \dots
\begin{diagram}
\draw (0.5,1.5) -- (1,1.5); 
\draw[rounded corners] (1,2) rectangle (2,1);
\draw (1.5,1.5) node (X) {$A$};
\draw (2,1.5) -- (3,1.5); 
\draw[rounded corners] (3,2) rectangle (4,1);
\draw (3.5,1.5) node {$A$};
\draw (4,1.5) -- (5,1.5);
\draw[rounded corners] (5,2) rectangle (6,1);
\draw (5.5,1.5) node {$A$};
\draw (6,1.5) -- (7,1.5); 
\draw[rounded corners] (7,2) rectangle (8,1);
\draw (7.5,1.5) node {$A$};
\draw (8,1.5) -- (9,1.5); 
\draw[rounded corners] (9,2) rectangle (10,1);
\draw (9.5,1.5) node {$A$};
\draw (10,1.5) -- (10.5,1.5);
\draw (1.5,1) -- (1.5,.5); \draw (3.5,1) -- (3.5,.5); \draw (5.5,1) -- (5.5,.5);
\draw (7.5,1) -- (7.5,.5); \draw (9.5,1) -- (9.5,.5); 
\end{diagram} \dots,
\end{equation}
i.e. the state is built up by concatenating different copies of the same tensor $A$. The last index of each tensor is contracted with the first index of the next one, whereas the uncontracted indices correspond to the physical degrees of freedom in the spin chain. Because the same tensor is repeated on every site in the chain the state is clearly translation invariant, and the contraction of the virtual indices allows for the state to exhibit strong quantum correlations. The (highly non-linear) manifold of matrix product states is defined by all states of the above form, and we can find an approximation of the ground state for a given Hamiltonian by variationally optimizing the tensor $A$. If we approximate all long-range interactions by a sum of exponentials -- an approximation that can be made arbitrarily precise, and which we will always perform in the following -- this optimization can be done efficiently using the algorithm in Ref. \cite{Zauner-Stauber2017}.
\par On this correlated background state, we can now build quasiparticle excitations by introducing a variational ansatz of the form \cite{Haegeman2012a}
\begin{equation} 
\ket{\Phi_k(B)} = \sum_{n} \e^{ikn}
\begin{diagram}
\draw (0.5,1.5) -- (1,1.5); 
\draw[rounded corners] (1,2) rectangle (2,1);
\draw (1.5,1.5) node (X) {$A$};
\draw (2,1.5) -- (3,1.5); 
\draw[rounded corners] (3,2) rectangle (4,1);
\draw (3.5,1.5) node {$A$};
\draw (4,1.5) -- (5,1.5);
\draw[rounded corners] (5,2) rectangle (6,1);
\draw (5.5,1.5) node {$B$};
\draw (6,1.5) -- (7,1.5); 
\draw[rounded corners] (7,2) rectangle (8,1);
\draw (7.5,1.5) node {$A$};
\draw (8,1.5) -- (9,1.5); 
\draw[rounded corners] (9,2) rectangle (10,1);
\draw (9.5,1.5) node {$A$};
\draw (10,1.5) -- (10.5,1.5);
\draw (1.5,1) -- (1.5,.5); \draw (3.5,1) -- (3.5,.5); \draw (5.5,1) -- (5.5,.5);
\draw (7.5,1) -- (7.5,.5); \draw (9.5,1) -- (9.5,.5);
\draw (1.5,0) node {$\dots$}; \draw (3.5,0) node {$s_{n-1}$}; \draw (5.5,0) node {$s_{n}$}; \draw (7.5,0) node {$s_{n+1}$}; \draw (9.5,0) node {$\dots$};
\end{diagram} \;.
\end{equation}
This ansatz is the momentum superposition of a local perturbation of the ground state. All variational degrees of freedom are contained in the tensor $B$, and, since it acts on the virtual degrees of freedom of the matrix product state, the tensor can use the correlations in the ground state to perturb the state over an extended region. In that way, it can describe a dressed quasiparticle -- a lump on the strongly-correlated background -- within a specific momentum sector. Since the ansatz state is linear in the tensor $B$ and it can be easily chosen to be orthogonal to the ground state, the variational optimization amounts to solving an eigenvalue problem. For both local and long-range interactions, this eigenvalue problem can be implemented and solved efficiently, so that we find the energies $\omega_i(k)$ and states $\ket{\Phi_k(B)}_i$ of the lowest-lying excitations within any momentum sector $k$.
\par This approach to describe dressed quasiparticles is orthogonal to the perturbative approach in, e.g., Fermi-liquid theory, where the quasiparticles are defined in a non-interacting limit and assumed to remain well-defined modes if the interactions are turned on. Indeed, we target the exact eigenstates for an interacting system directly, and, as such, describe stable quasiparticles -- decays can be treated by extending the formalism to multiparticle excitations \cite{Vanderstraeten2015a, Vanderstraeten2016}. For local Hamiltonians this approach leads to accurate results on the spectral properties of strongly-interacting spin chains \cite{Haegeman2012a, Vanderstraeten2015a, Vanderstraeten2016}, even in cases where the quasiparticle properties cannot be perturbatively connected to a non-interacting limit. In the following we will show that this approach can be extended to systems with long-range interactions.

\begin{figure}
\begin{center}
\includegraphics[width=\columnwidth]{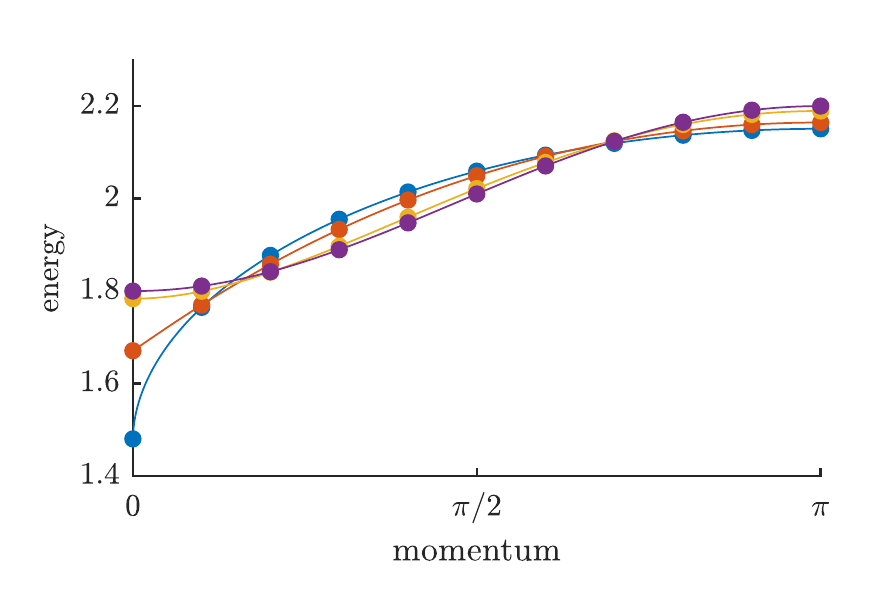}
	\caption{The dispersion relation of the extended long-range Ising model [Eq.~\eqref{eq:isinglike}] with the quasiparticle ansatz (markers) compared to the exact solution (lines) for $\lambda=10$ and a set of values for the long-range interaction: $\alpha=100$ (purple), $\alpha=4$ (yellow), $\alpha=2$ (red), and $\alpha=1.5$ (blue). In order to compare the dispersion relations on the same plot, we have rescaled the energy by the factor $\eta(\alpha,\lambda)=\sum_{n>1}n^{-\alpha}+\lambda$.}
		\label{fig:isinglike}
	\end{center}
\end{figure}

\noindent\emph{Benchmarking the ansatz}---
We will first test our quasiparticle ansatz on two benchmark models. The first is an extended long-range Ising model defined by the Hamiltonian
\begin{equation} \label{eq:isinglike}
H_{\text{ELI}} = - \sum_{i<j} \frac{1}{(j-i)^\alpha} \sigma^z_i \left( \prod_{i<n<j} \sigma^x_n \right) \sigma^z_j - \lambda \sum_i \sigma^x_i.
\end{equation}
Because of the string of $\sigma_x$ operators in the interaction term, this model can be mapped to a system of free electrons, for which the excitation spectrum can be computed exactly. In Fig.~\ref{fig:isinglike} we have plotted the dispersion relation as obtained by the quasiparticle ansatz, and compared to the exact solution. We see perfect agreement for different values of $\alpha$, and we can accurately resolve a cusp in the dispersion relation for $\alpha<2$.

\par
\begin{figure}
\begin{center}
\includegraphics[width=\columnwidth]{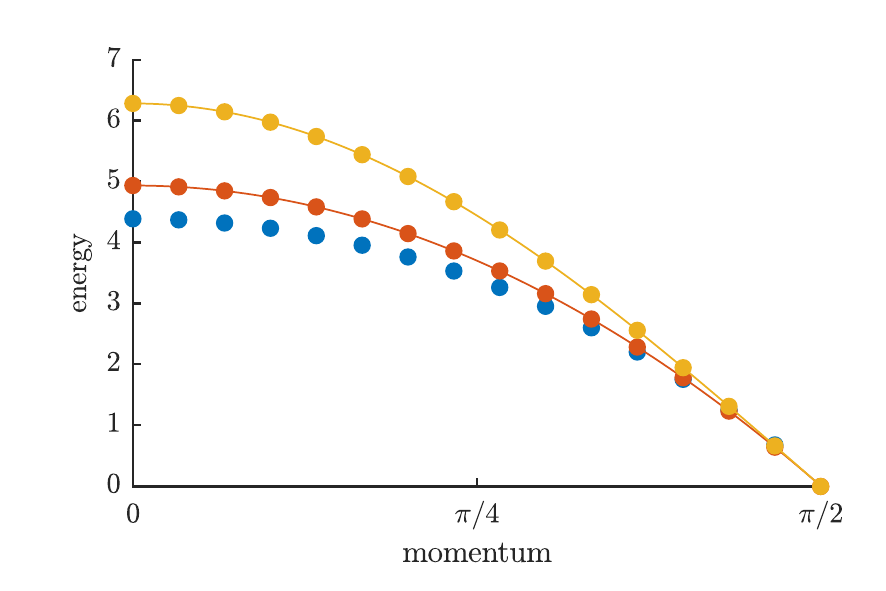}
	\caption{The spinon dispersion relation of the long-range Heisenberg model as computed with the quasiparticle ansatz for $\alpha=100$ (yellow), $\alpha=2$ (red) and $\alpha=1.5$ (blue). The full lines are the exact results for the nearest-neighbour model \cite{Faddeev1981} and the Haldane-Shastry model \cite{Bernevig2001}. The momentum of a single spinon in the thermodynamic limit is defined up to a constant shift, which we have fixed according to Ref.~\onlinecite{Bernevig2001}.}
    \label{fig:heisenberg}
	\end{center}
\end{figure}

 The second test case is the long-range Heisenberg model as defined by
\begin{equation}
H_{\text{Heis}} = \sum_{i<j} \frac{1}{(j-i)^\alpha} \left( \sigma^x_i \sigma^x_j + \sigma^y_i \sigma^y_j + \sigma^z_i \sigma^z_j \right).
\end{equation}
For $\alpha\rightarrow\infty$ this system reduces to the well-known Heisenberg model, whereas for $\alpha=2$ we recover the Haldane-Shastry model \cite{Haldane1988,Shastry1988}. For both cases the excitation spectrum can be determined exactly with the Bethe ansatz, but for all other $\alpha$ the model loses integrability. From the Bethe ansatz it is well known that the elementary excitations have a topological nature \cite{Faddeev1981} -- they are so-called spinons -- but we can easily extend our quasiparticle ansatz to also capture these topological excitations \cite{Haegeman2012a}. In Fig.~\ref{fig:heisenberg} we have plotted the spinon dispersion relation for three values of $\alpha$, and compared with the two exact solutions. Again, we see that we reproduce the dispersion relation accurately, and, more interestingly, there is no cusp for $\alpha<2$. This can be expected since the long-range interactions are frustrating in this case and, therefore, do not present a relevant perturbation on the nearest-neighbour interaction for all $\alpha>1$ \cite{Laflorencie2005, Gong2016}.

\noindent\emph{Long-range Ising model}--- %
Let us now study the prototypical long-range interacting spin chain, the transverse-field Ising model
\begin{equation} \label{eq:ising}
H_{\text{LTI}} = - \sum_{i<j} \frac{\sigma^z_i \sigma^z_j}{(j-i)^\alpha}  - \lambda \sum_i \sigma^x_i.
\end{equation}
In the nearest-neighbour limit ($\alpha\rightarrow\infty$) this model exhibits a second-order quantum phase transition at $\lambda_c=1$ from a ferromagnetic ($\lambda<1$) to a paramagnetic ($\lambda>1$) state \cite{Sachdev2011}. This phase transition persists if the long-range interactions are turned on, but the critical value $\lambda_c(\alpha)$ shifts to higher values. Deep in the paramagnetic phase, we expect that linear spin-wave theory yields a good approximation of the dynamical properties \cite{Hauke2013a, Cevolani2016, Cevolani2017}, but with the quasiparticle ansatz we can study to what extent this spin-wave picture continues to hold when approaching the critical point. Although the phase diagram remains qualitatively the same, the long-range interactions induce correlation functions with power-law decay in the ground state even away from criticality \cite{Deng2005}.
\par
\begin{figure}
\includegraphics[width=\columnwidth]{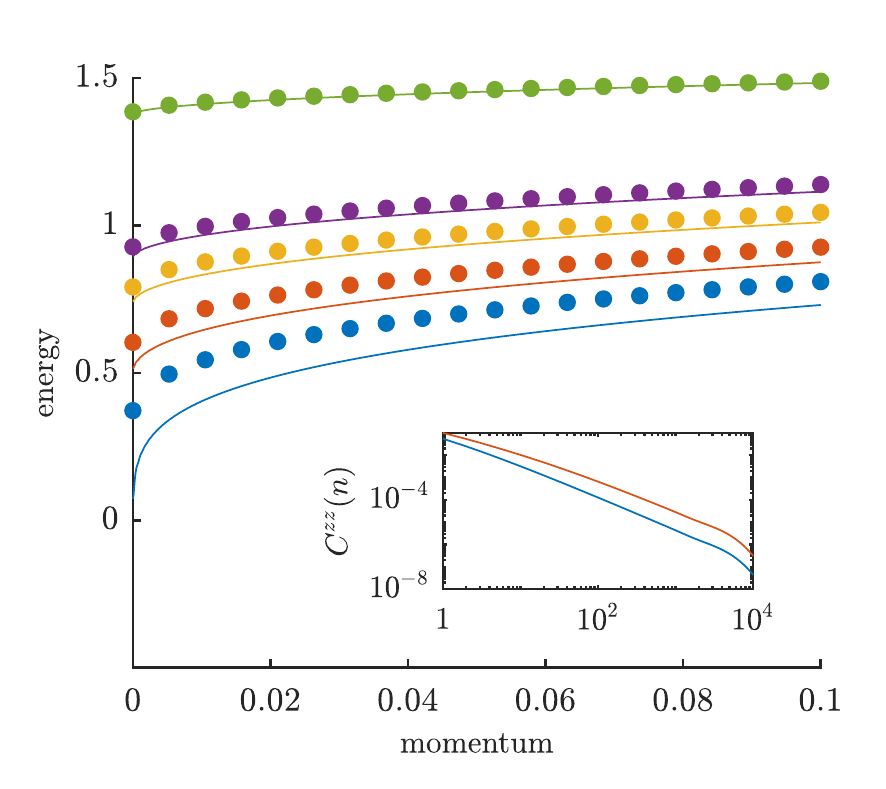}
\caption{The dispersion relation for the long-range Ising model [Eq.~\eqref{eq:ising}] computed with the quasiparticle ansatz (markers), and compared with spin-wave theory (lines) for $\alpha=1.5$ and $\lambda=5.2$ (blue), $\lambda=6$ (red), $\lambda=7$ (yellow), $\lambda=8$ (purple) and $\lambda=15$ (green). Again, the energy has been rescaled by the factor $\eta(\alpha,\lambda)$. For reference, the inset contains the ground-state correlation function $C^{zz}(n)=\braket{\sigma^z_n\sigma^z_0}$ for $\alpha=1.5$, $\lambda=10$ (blue) and $\lambda=6$ (red), showing power-law decay $C^{zz}(n)\propto n^{-\gamma}$ with powers $\gamma\approx1.47$ and $\gamma\approx1.37$, respectively; we obtain significant deviations form the spin-wave result (for which $\gamma$ equals $\alpha$ \cite{Deng2005}), in agreement with Ref.~\cite{Vodola2016}. For very long distances, the MPS approximation induces an exponential decay.}
\label{fig:disp}
\end{figure}

 In Fig.~\ref{fig:disp} our results for the Ising model are presented. In the inset we show that the ground-state correlations in our MPS simulation accurately reproduce the correct power-law decay. Building on this ground state, our quasiparticle ansatz reproduces the cusp in the dispersion relation for small momenta (for higher momenta the dispersion flattens out further). We observe that for large $\lambda$ our results coincide with spin-wave theory, but the latter induces significant errors when lowering $\lambda$. In particular, using the closing of the gap as a criterion, we find different values for the critical point; e.g., we find a closing of the gap at $\lambda_c\approx4.70$ for $\alpha=1.5$, whereas spin-wave theory predicts the considerably larger value $\lambda_c=5.22$. 
\par
\begin{figure}
\includegraphics[width=\columnwidth]{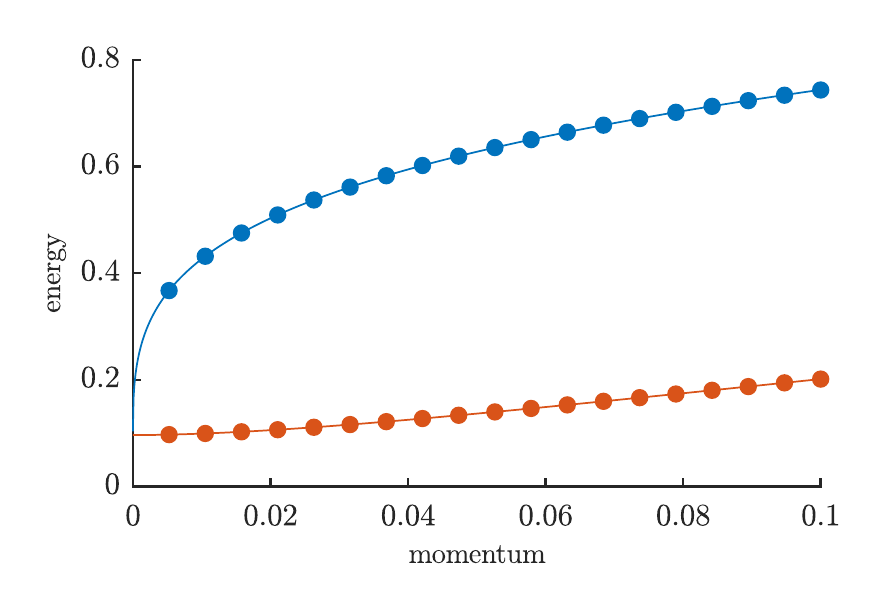}
\caption{A close-up of the dispersion relation of the Ising model [Eq.~\eqref{eq:ising}] computed with the quasiparticle ansatz (markers) for two values of $(\alpha,\lambda)$ close to the critical point: $(1.5,4.8)$ (blue) and $(2.875,1.586)$ (red). The lines are fits to the universal long-wavelength form in Eq.~\eqref{eq:universal}. Again, the energy has been rescaled by the factor $\eta(\alpha,\lambda)$.}
\label{fig:cusp}
\end{figure}

 In order to study the cusp in more detail we have plotted a close-up for two values of $\alpha$ and a corresponding values of $\lambda$ close to the critical value $\lambda_c(\alpha)$. Spin-wave theory suggests that we should find a dispersion relation in the long-wavelength limit
\begin{equation} \label{eq:universal}
\omega(k) = \sqrt{\Delta^2 + a k^{\alpha-1} + b k^2}, \qquad k\rightarrow 0
\end{equation}
where $\Delta$ is the gap. As we have seen in Fig.~\ref{fig:disp} the gap $\Delta$ is shifted considerably from the spin-wave result, but, as we can see from Fig.~\ref{fig:cusp}, the universal form of the cusp in the dispersion relation remains correct. Therefore, we expect that the cusp's signature in the quench dynamics \cite{Maghrebi2016, Cevolani2017} is visible deep in the interacting regime as well.

\par
\begin{figure}
\includegraphics[width=.95\columnwidth]{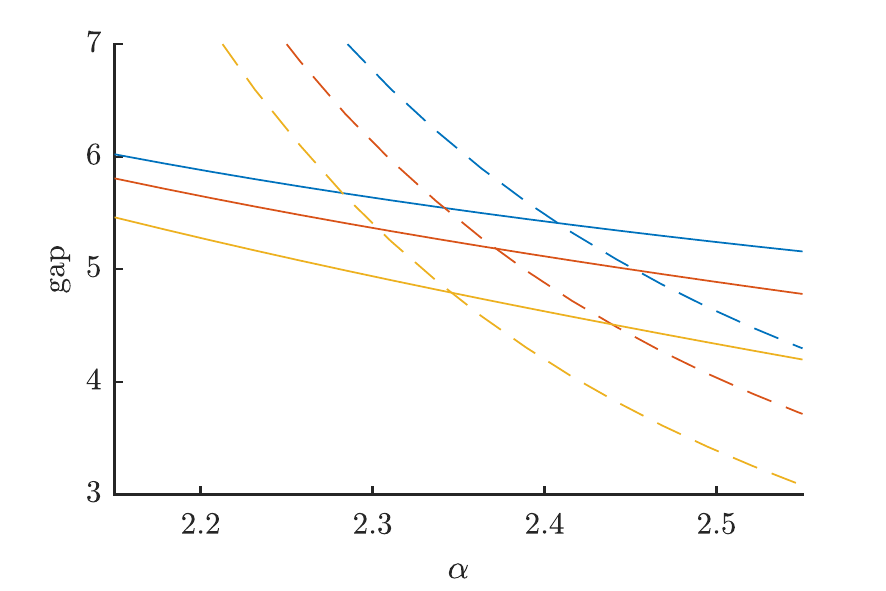}
\caption{The gap for the `topological' excitation (striped) and for the `trivial' excitation (full) as a function of $\alpha$ for three different values of the magnetic field: $\lambda=0.25$ (blue), $\lambda=0.5$ (red), and $\lambda=0.75$ (yellow).}
\label{fig:domain}
\end{figure}

 On the other side of the phase transition, in the symmetry-broken phase, the elementary excitations can have a topological nature. Indeed, in the nearest-neighbour limit it is known that the low-energy quasiparticles are domain walls between the two symmetry-broken ground-state configurations \cite{Sachdev2011}. Upon lowering $\alpha$, however, the long-range interactions will induce an increasing energy cost associated to the two ground-state configurations across the domain wall. Therefore we expect that another non-topological quasiparticle excitation will, for small enough $\alpha$, have lower energy and dominate the low-energy behavior of the system, whereas the domain wall persists as a stable but heavier quasiparticle. In Fig.~\ref{fig:domain} we have plotted the gap of both the domain wall and the local excitation as a function of $\alpha$ for three values of the magnetic field, showing that the crossover between the two gaps occurs around $\alpha\approx2.3-2.4$. We expect this crossover to have drastic effects on the quench dynamics of this model, as these dynamics are assumed to be determined by the spreading of the lowest-lying quasiparticles \cite{Calabrese2006, Calabrese2007a}. For example, the value of $\alpha$ roughly coincides with the region where an anomalous dynamical phase transition occurs \cite{Halimeh2017, Homrighausen2017}, which suggests that this crossover between topological and trivial quasiparticles provides the physical origin of this phenomenon.

\begin{figure}
\begin{center}
\includegraphics[width=\columnwidth]{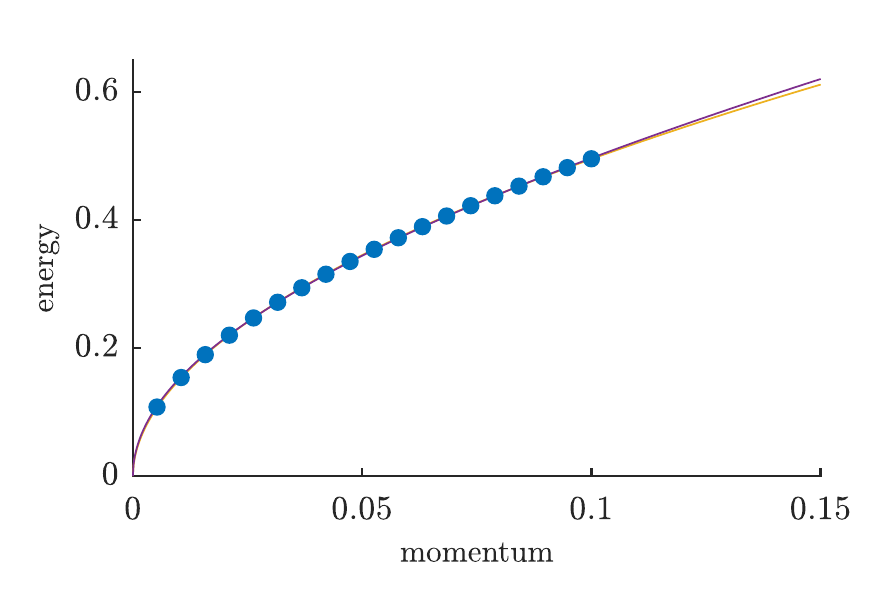}
	\caption{Dispersion relation for the XXZ model [Eq.\eqref{eq:xxz}] with $\Delta=0$ and $\alpha=2$ with the quasiparticle ansatz, and compared to the form  $\omega(k) = (ak^{\alpha-1}+bk^2 )^{1/2}$ (purple) and the form $\omega(k)=az^k$ (yellow). We have extended the range of the figure in order to see the deviation of the two fits.}
		\label{fig:xx}
	\end{center}
\end{figure}

\noindent\emph{XXZ model}--- %
Finally we study the XXZ model as defined by the Hamiltonian
\begin{equation} \label{eq:xxz}
H_{\text{XXZ}} = \sum_{i<j} \frac{1}{(j-i)^\alpha} \left( - \sigma^x_i \sigma^x_j - \sigma^y_i \sigma^y_j + \Delta \sigma^z_i \sigma^z_j \right).
\end{equation}
In the region $|\Delta|<1$ and for large $\alpha$ the model is in the Luttinger-liquid phase, whereas the continuous $U(1)$ symmetry is broken for small enough $\alpha$ \cite{Maghrebi2017}. Although this transition is hard to see in standard MPS simulations, it is nicely visible in the spectrum: for the Luttinger liquid the excitations have a linear dispersion ($z=1$), whereas in the symmetry-broken phase the spectrum consists of Goldstone modes with a cusp in the dispersion relation ($z<1$). In Fig.~\ref{fig:xx} we have plotted our results for the dispersion relation at $\alpha=2$ (in the symmetry-broken phase), and fitted this to the form in Eq.~\eqref{eq:universal} with $\Delta=0$. In the long-wavelength limit only the term $\omega(k)\propto k^{(\alpha-1)/2}$ is expected to survive, but this appears to be a poor fit on the momentum range that we have considered in Fig.~\ref{fig:xx}. Instead, it appears that we could nicely fit a pure power law $\omega(k)\propto k^z$ to the numerical data with the exponent $z=0.5205(11)$. This allows us to define an effective dynamic critical exponent, which captures the cusp in the dispersion relation for a significant portion of the Brillouin zone, and which is expected to determine the correlation spreading in this system.

\noindent\emph{Conclusions}--- %
In this paper we have computed the dispersion relations of a number of gapped and gapless quantum spin chains with long-range interactions using a variational ansatz with an explicit quasiparticle nature. The accuracy of these computations shows that quasiparticles continue to capture the low-energy degrees of freedom in strongly-correlated regimes, and provide the key to understanding all low-energy dynamical properties. In particular, we have shown that these quasiparticles can exhibit unexpected properties -- cusps in the quasiparticle dispersion, the crossover between trivial and topological particles and the emergence of an effective dynamical critical exponent are points in case.

\noindent\emph{Acknowledgements}--- %
We acknowledge inspiring discussions with Valentin Zauner-Stauber, Jutho Haegeman and Jad Halimeh. This work was supported by the Flemish Research Foundation, the Austrian Science Fund (ViCoM, FoQuS), the European Commission (QUTE 647905, ERQUAF 715861), and the Deutsche Forschungsgemeinschaft (DFG) within the research unit FOR 2247.

\bibliography{./bibliography}

\begin{thebibliography}{40}%
\makeatletter
\providecommand \@ifxundefined [1]{%
 \@ifx{#1\undefined}
}%
\providecommand \@ifnum [1]{%
 \ifnum #1\expandafter \@firstoftwo
 \else \expandafter \@secondoftwo
 \fi
}%
\providecommand \@ifx [1]{%
 \ifx #1\expandafter \@firstoftwo
 \else \expandafter \@secondoftwo
 \fi
}%
\providecommand \natexlab [1]{#1}%
\providecommand \enquote  [1]{``#1''}%
\providecommand \bibnamefont  [1]{#1}%
\providecommand \bibfnamefont [1]{#1}%
\providecommand \citenamefont [1]{#1}%
\providecommand \href@noop [0]{\@secondoftwo}%
\providecommand \href [0]{\begingroup \@sanitize@url \@href}%
\providecommand \@href[1]{\@@startlink{#1}\@@href}%
\providecommand \@@href[1]{\endgroup#1\@@endlink}%
\providecommand \@sanitize@url [0]{\catcode `\\12\catcode `\$12\catcode
  `\&12\catcode `\#12\catcode `\^12\catcode `\_12\catcode `\%12\relax}%
\providecommand \@@startlink[1]{}%
\providecommand \@@endlink[0]{}%
\providecommand \url  [0]{\begingroup\@sanitize@url \@url }%
\providecommand \@url [1]{\endgroup\@href {#1}{\urlprefix }}%
\providecommand \urlprefix  [0]{URL }%
\providecommand \Eprint [0]{\href }%
\providecommand \doibase [0]{http://dx.doi.org/}%
\providecommand \selectlanguage [0]{\@gobble}%
\providecommand \bibinfo  [0]{\@secondoftwo}%
\providecommand \bibfield  [0]{\@secondoftwo}%
\providecommand \translation [1]{[#1]}%
\providecommand \BibitemOpen [0]{}%
\providecommand \bibitemStop [0]{}%
\providecommand \bibitemNoStop [0]{.\EOS\space}%
\providecommand \EOS [0]{\spacefactor3000\relax}%
\providecommand \BibitemShut  [1]{\csname bibitem#1\endcsname}%
\let\auto@bib@innerbib\@empty
\bibitem [{\citenamefont {Schau{\ss}}\ \emph {et~al.}(2012)\citenamefont
  {Schau{\ss}}, \citenamefont {Cheneau}, \citenamefont {Endres}, \citenamefont
  {Fukuhara}, \citenamefont {Hild}, \citenamefont {Omran}, \citenamefont
  {Pohl}, \citenamefont {Gross}, \citenamefont {Kuhr},\ and\ \citenamefont
  {Bloch}}]{Schauss2012}%
  \BibitemOpen
  \bibfield  {author} {\bibinfo {author} {\bibfnamefont {P.}~\bibnamefont
  {Schau{\ss}}}, \bibinfo {author} {\bibfnamefont {M.}~\bibnamefont {Cheneau}},
  \bibinfo {author} {\bibfnamefont {M.}~\bibnamefont {Endres}}, \bibinfo
  {author} {\bibfnamefont {T.}~\bibnamefont {Fukuhara}}, \bibinfo {author}
  {\bibfnamefont {S.}~\bibnamefont {Hild}}, \bibinfo {author} {\bibfnamefont
  {A.}~\bibnamefont {Omran}}, \bibinfo {author} {\bibfnamefont
  {T.}~\bibnamefont {Pohl}}, \bibinfo {author} {\bibfnamefont {C.}~\bibnamefont
  {Gross}}, \bibinfo {author} {\bibfnamefont {S.}~\bibnamefont {Kuhr}}, \ and\
  \bibinfo {author} {\bibfnamefont {I.}~\bibnamefont {Bloch}},\ }\bibfield
  {title} {\enquote {\bibinfo {title} {Observation of spatially ordered
  structures in a two-dimensional rydberg gas},}\ }\href
  {https://www.nature.com/articles/nature11596} {\bibfield  {journal} {\bibinfo
   {journal} {Nature}\ }\textbf {\bibinfo {volume} {491}},\ \bibinfo {pages}
  {87--91} (\bibinfo {year} {2012})}\BibitemShut {NoStop}%
\bibitem [{\citenamefont {Peter}\ \emph {et~al.}(2012)\citenamefont {Peter},
  \citenamefont {M{\"{u}}ller}, \citenamefont {Wessel},\ and\ \citenamefont
  {B{\"{u}}chler}}]{Peter2012}%
  \BibitemOpen
  \bibfield  {author} {\bibinfo {author} {\bibfnamefont {D.}~\bibnamefont
  {Peter}}, \bibinfo {author} {\bibfnamefont {S.}~\bibnamefont {M{\"{u}}ller}},
  \bibinfo {author} {\bibfnamefont {S.}~\bibnamefont {Wessel}}, \ and\ \bibinfo
  {author} {\bibfnamefont {H.~P.}\ \bibnamefont {B{\"{u}}chler}},\ }\bibfield
  {title} {\enquote {\bibinfo {title} {{Anomalous Behavior of Spin Systems with
  Dipolar Interactions}},}\ }\href {\doibase 10.1103/PhysRevLett.109.025303}
  {\bibfield  {journal} {\bibinfo  {journal} {Physical Review Letters}\
  }\textbf {\bibinfo {volume} {109}},\ \bibinfo {pages} {025303} (\bibinfo
  {year} {2012})}\BibitemShut {NoStop}%
\bibitem [{\citenamefont {Richerme}\ \emph {et~al.}(2014)\citenamefont
  {Richerme}, \citenamefont {Gong}, \citenamefont {Lee}, \citenamefont {Senko},
  \citenamefont {Smith}, \citenamefont {Foss-Feig}, \citenamefont {Michalakis},
  \citenamefont {Gorshkov},\ and\ \citenamefont {Monroe}}]{Richerme2014}%
  \BibitemOpen
  \bibfield  {author} {\bibinfo {author} {\bibfnamefont {P.}~\bibnamefont
  {Richerme}}, \bibinfo {author} {\bibfnamefont {Z.-X.}\ \bibnamefont {Gong}},
  \bibinfo {author} {\bibfnamefont {A.}~\bibnamefont {Lee}}, \bibinfo {author}
  {\bibfnamefont {C.}~\bibnamefont {Senko}}, \bibinfo {author} {\bibfnamefont
  {J.}~\bibnamefont {Smith}}, \bibinfo {author} {\bibfnamefont
  {M.}~\bibnamefont {Foss-Feig}}, \bibinfo {author} {\bibfnamefont
  {S.}~\bibnamefont {Michalakis}}, \bibinfo {author} {\bibfnamefont {A.~V.}\
  \bibnamefont {Gorshkov}}, \ and\ \bibinfo {author} {\bibfnamefont
  {C.}~\bibnamefont {Monroe}},\ }\bibfield  {title} {\enquote {\bibinfo {title}
  {Non-local propagation of correlations in quantum systems with long-range
  interactions},}\ }\href {https://www.nature.com/articles/nature13450}
  {\bibfield  {journal} {\bibinfo  {journal} {Nature}\ }\textbf {\bibinfo
  {volume} {511}},\ \bibinfo {pages} {198--201} (\bibinfo {year}
  {2014})}\BibitemShut {NoStop}%
\bibitem [{\citenamefont {Jurcevic}\ \emph {et~al.}(2014)\citenamefont
  {Jurcevic}, \citenamefont {Lanyon}, \citenamefont {Hauke}, \citenamefont
  {Hempel}, \citenamefont {Zoller}, \citenamefont {Blatt},\ and\ \citenamefont
  {Roos}}]{Jurcevic2014}%
  \BibitemOpen
  \bibfield  {author} {\bibinfo {author} {\bibfnamefont {P.}~\bibnamefont
  {Jurcevic}}, \bibinfo {author} {\bibfnamefont {B.~P.}\ \bibnamefont
  {Lanyon}}, \bibinfo {author} {\bibfnamefont {P.}~\bibnamefont {Hauke}},
  \bibinfo {author} {\bibfnamefont {C.}~\bibnamefont {Hempel}}, \bibinfo
  {author} {\bibfnamefont {P.}~\bibnamefont {Zoller}}, \bibinfo {author}
  {\bibfnamefont {R.}~\bibnamefont {Blatt}}, \ and\ \bibinfo {author}
  {\bibfnamefont {C.~F.}\ \bibnamefont {Roos}},\ }\bibfield  {title} {\enquote
  {\bibinfo {title} {Quasiparticle engineering and entanglement propagation in
  a quantum many-body system},}\ }\href
  {https://www.nature.com/articles/nature13461} {\bibfield  {journal} {\bibinfo
   {journal} {Nature}\ }\textbf {\bibinfo {volume} {511}},\ \bibinfo {pages}
  {202--205} (\bibinfo {year} {2014})}\BibitemShut {NoStop}%
\bibitem [{\citenamefont {Mermin}\ and\ \citenamefont
  {Wagner}(1966)}]{Mermin1966}%
  \BibitemOpen
  \bibfield  {author} {\bibinfo {author} {\bibfnamefont {N.~D.}\ \bibnamefont
  {Mermin}}\ and\ \bibinfo {author} {\bibfnamefont {H.}~\bibnamefont
  {Wagner}},\ }\bibfield  {title} {\enquote {\bibinfo {title} {{Absence of
  Ferromagnetism or Antiferromagnetism in One- or Two-Dimensional Isotropic
  Heisenberg Models}},}\ }\href
  {http://link.aps.org/doi/10.1103/PhysRevLett.17.1133} {\bibfield  {journal}
  {\bibinfo  {journal} {Physical Review Letters}\ }\textbf {\bibinfo {volume}
  {17}},\ \bibinfo {pages} {1133--1136} (\bibinfo {year} {1966})}\BibitemShut
  {NoStop}%
\bibitem [{\citenamefont {Maghrebi}\ \emph {et~al.}(2017)\citenamefont
  {Maghrebi}, \citenamefont {Gong},\ and\ \citenamefont
  {Gorshkov}}]{Maghrebi2017}%
  \BibitemOpen
  \bibfield  {author} {\bibinfo {author} {\bibfnamefont {M.~F.}\ \bibnamefont
  {Maghrebi}}, \bibinfo {author} {\bibfnamefont {Z.-X.}\ \bibnamefont {Gong}},
  \ and\ \bibinfo {author} {\bibfnamefont {A.~V.}\ \bibnamefont {Gorshkov}},\
  }\bibfield  {title} {\enquote {\bibinfo {title} {Continuous symmetry breaking
  in 1d long-range interacting quantum systems},}\ }\href {\doibase
  10.1103/PhysRevLett.119.023001} {\bibfield  {journal} {\bibinfo  {journal}
  {Physical Review Letters}\ }\textbf {\bibinfo {volume} {119}},\ \bibinfo
  {pages} {023001} (\bibinfo {year} {2017})}\BibitemShut {NoStop}%
\bibitem [{\citenamefont {Lieb}\ and\ \citenamefont
  {Robinson}(1972)}]{Lieb1972}%
  \BibitemOpen
  \bibfield  {author} {\bibinfo {author} {\bibfnamefont {E.~H.}\ \bibnamefont
  {Lieb}}\ and\ \bibinfo {author} {\bibfnamefont {D.~W.}\ \bibnamefont
  {Robinson}},\ }\bibfield  {title} {\enquote {\bibinfo {title} {{The finite
  group velocity of quantum spin systems}},}\ }\href {\doibase
  10.1007/BF01645779} {\bibfield  {journal} {\bibinfo  {journal}
  {Communications in Mathematical Physics}\ }\textbf {\bibinfo {volume} {28}},\
  \bibinfo {pages} {251--257} (\bibinfo {year} {1972})}\BibitemShut {NoStop}%
\bibitem [{\citenamefont {Hastings}(2007)}]{Hastings2007}%
  \BibitemOpen
  \bibfield  {author} {\bibinfo {author} {\bibfnamefont {M.~B.}\ \bibnamefont
  {Hastings}},\ }\bibfield  {title} {\enquote {\bibinfo {title} {{An area law
  for one-dimensional quantum systems}},}\ }\href {\doibase
  10.1088/1742-5468/2007/08/P08024} {\bibfield  {journal} {\bibinfo  {journal}
  {Journal of Statistical Mechanics: Theory and Experiment}\ }\textbf {\bibinfo
  {volume} {2007}},\ \bibinfo {pages} {P08024} (\bibinfo {year}
  {2007})}\BibitemShut {NoStop}%
\bibitem [{\citenamefont {Hastings}\ and\ \citenamefont
  {Koma}(2006)}]{Hastings2006a}%
  \BibitemOpen
  \bibfield  {author} {\bibinfo {author} {\bibfnamefont {M.~B.}\ \bibnamefont
  {Hastings}}\ and\ \bibinfo {author} {\bibfnamefont {T.}~\bibnamefont
  {Koma}},\ }\bibfield  {title} {\enquote {\bibinfo {title} {{Spectral Gap and
  Exponential Decay of Correlations}},}\ }\href {\doibase
  10.1007/s00220-006-0030-4} {\bibfield  {journal} {\bibinfo  {journal}
  {Communications in Mathematical Physics}\ }\textbf {\bibinfo {volume}
  {265}},\ \bibinfo {pages} {781--804} (\bibinfo {year} {2006})}\BibitemShut
  {NoStop}%
\bibitem [{\citenamefont {Foss-Feig}\ \emph {et~al.}(2015)\citenamefont
  {Foss-Feig}, \citenamefont {Gong}, \citenamefont {Clark},\ and\ \citenamefont
  {Gorshkov}}]{Foss-Feig2015}%
  \BibitemOpen
  \bibfield  {author} {\bibinfo {author} {\bibfnamefont {M.}~\bibnamefont
  {Foss-Feig}}, \bibinfo {author} {\bibfnamefont {Z.-X.}\ \bibnamefont {Gong}},
  \bibinfo {author} {\bibfnamefont {C.~W.}\ \bibnamefont {Clark}}, \ and\
  \bibinfo {author} {\bibfnamefont {A.~V.}\ \bibnamefont {Gorshkov}},\
  }\bibfield  {title} {\enquote {\bibinfo {title} {Nearly linear light cones in
  long-range interacting quantum systems},}\ }\href {\doibase
  10.1103/PhysRevLett.114.157201} {\bibfield  {journal} {\bibinfo  {journal}
  {Physical Review Letters}\ }\textbf {\bibinfo {volume} {114}},\ \bibinfo
  {pages} {157201} (\bibinfo {year} {2015})}\BibitemShut {NoStop}%
\bibitem [{\citenamefont {Hauke}\ and\ \citenamefont
  {Tagliacozzo}(2013)}]{Hauke2013a}%
  \BibitemOpen
  \bibfield  {author} {\bibinfo {author} {\bibfnamefont {P.}~\bibnamefont
  {Hauke}}\ and\ \bibinfo {author} {\bibfnamefont {L.}~\bibnamefont
  {Tagliacozzo}},\ }\bibfield  {title} {\enquote {\bibinfo {title} {{Spread of
  Correlations in Long-Range Interacting Quantum Systems.}}}\ }\href {\doibase
  10.1103/PhysRevLett.111.207202} {\bibfield  {journal} {\bibinfo  {journal}
  {Physical Review Letters}\ }\textbf {\bibinfo {volume} {111}},\ \bibinfo
  {pages} {207202} (\bibinfo {year} {2013})}\BibitemShut {NoStop}%
\bibitem [{\citenamefont {Maghrebi}\ \emph {et~al.}(2016)\citenamefont
  {Maghrebi}, \citenamefont {Gong}, \citenamefont {Foss-Feig},\ and\
  \citenamefont {Gorshkov}}]{Maghrebi2016}%
  \BibitemOpen
  \bibfield  {author} {\bibinfo {author} {\bibfnamefont {M.~F.}\ \bibnamefont
  {Maghrebi}}, \bibinfo {author} {\bibfnamefont {Z.-X.}\ \bibnamefont {Gong}},
  \bibinfo {author} {\bibfnamefont {M.}~\bibnamefont {Foss-Feig}}, \ and\
  \bibinfo {author} {\bibfnamefont {A.~V.}\ \bibnamefont {Gorshkov}},\
  }\bibfield  {title} {\enquote {\bibinfo {title} {Causality and quantum
  criticality in long-range lattice models},}\ }\href {\doibase
  10.1103/PhysRevB.93.125128} {\bibfield  {journal} {\bibinfo  {journal}
  {Physical Review B}\ }\textbf {\bibinfo {volume} {93}},\ \bibinfo {pages}
  {125128} (\bibinfo {year} {2016})}\BibitemShut {NoStop}%
\bibitem [{\citenamefont {Buyskikh}\ \emph {et~al.}(2016)\citenamefont
  {Buyskikh}, \citenamefont {Fagotti}, \citenamefont {Schachenmayer},
  \citenamefont {Essler},\ and\ \citenamefont {Daley}}]{Buyskikh2016}%
  \BibitemOpen
  \bibfield  {author} {\bibinfo {author} {\bibfnamefont {A.~S.}\ \bibnamefont
  {Buyskikh}}, \bibinfo {author} {\bibfnamefont {M.}~\bibnamefont {Fagotti}},
  \bibinfo {author} {\bibfnamefont {J.}~\bibnamefont {Schachenmayer}}, \bibinfo
  {author} {\bibfnamefont {F.}~\bibnamefont {Essler}}, \ and\ \bibinfo {author}
  {\bibfnamefont {A.~J.}\ \bibnamefont {Daley}},\ }\bibfield  {title} {\enquote
  {\bibinfo {title} {Entanglement growth and correlation spreading with
  variable-range interactions in spin and fermionic tunneling models},}\ }\href
  {\doibase 10.1103/PhysRevA.93.053620} {\bibfield  {journal} {\bibinfo
  {journal} {Physical Review A}\ }\textbf {\bibinfo {volume} {93}},\ \bibinfo
  {pages} {053620} (\bibinfo {year} {2016})}\BibitemShut {NoStop}%
\bibitem [{\citenamefont {Haegeman}\ \emph {et~al.}(2013)\citenamefont
  {Haegeman}, \citenamefont {Michalakis}, \citenamefont {Nachtergaele},
  \citenamefont {Osborne}, \citenamefont {Schuch},\ and\ \citenamefont
  {Verstraete}}]{Haegeman2013}%
  \BibitemOpen
  \bibfield  {author} {\bibinfo {author} {\bibfnamefont {J.}~\bibnamefont
  {Haegeman}}, \bibinfo {author} {\bibfnamefont {S.}~\bibnamefont
  {Michalakis}}, \bibinfo {author} {\bibfnamefont {B.}~\bibnamefont
  {Nachtergaele}}, \bibinfo {author} {\bibfnamefont {T.~J.}\ \bibnamefont
  {Osborne}}, \bibinfo {author} {\bibfnamefont {N.}~\bibnamefont {Schuch}}, \
  and\ \bibinfo {author} {\bibfnamefont {F.}~\bibnamefont {Verstraete}},\
  }\bibfield  {title} {\enquote {\bibinfo {title} {Elementary excitations in
  gapped quantum spin systems},}\ }\href {\doibase
  10.1103/PhysRevLett.111.080401} {\bibfield  {journal} {\bibinfo  {journal}
  {Physical Review Letters}\ }\textbf {\bibinfo {volume} {111}},\ \bibinfo
  {pages} {080401} (\bibinfo {year} {2013})}\BibitemShut {NoStop}%
\bibitem [{\citenamefont {Calabrese}\ and\ \citenamefont
  {Cardy}(2006)}]{Calabrese2006}%
  \BibitemOpen
  \bibfield  {author} {\bibinfo {author} {\bibfnamefont {P.}~\bibnamefont
  {Calabrese}}\ and\ \bibinfo {author} {\bibfnamefont {J.}~\bibnamefont
  {Cardy}},\ }\bibfield  {title} {\enquote {\bibinfo {title} {{Time Dependence
  of Correlation Functions Following a Quantum Quench}},}\ }\href {\doibase
  10.1103/PhysRevLett.96.136801} {\bibfield  {journal} {\bibinfo  {journal}
  {Physical Review Letters}\ }\textbf {\bibinfo {volume} {96}},\ \bibinfo
  {pages} {136801} (\bibinfo {year} {2006})}\BibitemShut {NoStop}%
\bibitem [{\citenamefont {Calabrese}\ and\ \citenamefont
  {Cardy}(2007)}]{Calabrese2007a}%
  \BibitemOpen
  \bibfield  {author} {\bibinfo {author} {\bibfnamefont {P.}~\bibnamefont
  {Calabrese}}\ and\ \bibinfo {author} {\bibfnamefont {J.}~\bibnamefont
  {Cardy}},\ }\bibfield  {title} {\enquote {\bibinfo {title} {{Quantum quenches
  in extended systems}},}\ }\href {\doibase 10.1088/1742-5468/2007/06/P06008}
  {\bibfield  {journal} {\bibinfo  {journal} {Journal of Statistical Mechanics:
  Theory and Experiment}\ }\textbf {\bibinfo {volume} {2007}},\ \bibinfo
  {pages} {P06008--P06008} (\bibinfo {year} {2007})}\BibitemShut {NoStop}%
\bibitem [{\citenamefont {Fey}\ and\ \citenamefont {Schmidt}(2016)}]{Fey2016}%
  \BibitemOpen
  \bibfield  {author} {\bibinfo {author} {\bibfnamefont {Sebastian}\
  \bibnamefont {Fey}}\ and\ \bibinfo {author} {\bibfnamefont {Kai~Phillip}\
  \bibnamefont {Schmidt}},\ }\bibfield  {title} {\enquote {\bibinfo {title}
  {Critical behavior of quantum magnets with long-range interactions in the
  thermodynamic limit},}\ }\href {\doibase 10.1103/PhysRevB.94.075156}
  {\bibfield  {journal} {\bibinfo  {journal} {Physical Review B}\ }\textbf
  {\bibinfo {volume} {94}},\ \bibinfo {pages} {075156} (\bibinfo {year}
  {2016})}\BibitemShut {NoStop}%
\bibitem [{\citenamefont {Cevolani}\ \emph {et~al.}(2016)\citenamefont
  {Cevolani}, \citenamefont {Carleo},\ and\ \citenamefont
  {Sanchez-Palencia}}]{Cevolani2016}%
  \BibitemOpen
  \bibfield  {author} {\bibinfo {author} {\bibfnamefont {L.}~\bibnamefont
  {Cevolani}}, \bibinfo {author} {\bibfnamefont {G.}~\bibnamefont {Carleo}}, \
  and\ \bibinfo {author} {\bibfnamefont {L.}~\bibnamefont {Sanchez-Palencia}},\
  }\bibfield  {title} {\enquote {\bibinfo {title} {Spreading of correlations in
  exactly solvable quantum models with long-range interactions in arbitrary
  dimensions},}\ }\href
  {http://iopscience.iop.org/article/10.1088/1367-2630/18/9/093002/meta}
  {\bibfield  {journal} {\bibinfo  {journal} {New Journal of Physics}\ }\textbf
  {\bibinfo {volume} {18}},\ \bibinfo {pages} {093002} (\bibinfo {year}
  {2016})}\BibitemShut {NoStop}%
\bibitem [{\citenamefont {Cevolani}\ \emph {et~al.}(2017)\citenamefont
  {Cevolani}, \citenamefont {Despres}, \citenamefont {Carleo}, \citenamefont
  {Tagliacozzo},\ and\ \citenamefont {Sanchez-Palencia}}]{Cevolani2017}%
  \BibitemOpen
  \bibfield  {author} {\bibinfo {author} {\bibfnamefont {L.}~\bibnamefont
  {Cevolani}}, \bibinfo {author} {\bibfnamefont {J.}~\bibnamefont {Despres}},
  \bibinfo {author} {\bibfnamefont {G.}~\bibnamefont {Carleo}}, \bibinfo
  {author} {\bibfnamefont {L.}~\bibnamefont {Tagliacozzo}}, \ and\ \bibinfo
  {author} {\bibfnamefont {L.}~\bibnamefont {Sanchez-Palencia}},\ }\bibfield
  {title} {\enquote {\bibinfo {title} {{Universal Scaling Laws for Correlation
  Spreading in Quantum Systems with Short- and Long-Range Interactions}},}\
  }\href {https://arxiv.org/abs/1706.00838} {\  (\bibinfo {year} {2017})},\
  \Eprint {http://arxiv.org/abs/1706.00838} {arXiv:1706.00838} \BibitemShut
  {NoStop}%
\bibitem [{\citenamefont {Fr\'erot}\ \emph {et~al.}(2018)\citenamefont
  {Fr\'erot}, \citenamefont {Naldesi},\ and\ \citenamefont
  {Roscilde}}]{Frerot2018}%
  \BibitemOpen
  \bibfield  {author} {\bibinfo {author} {\bibfnamefont {Ir\'en\'ee}\
  \bibnamefont {Fr\'erot}}, \bibinfo {author} {\bibfnamefont {Piero}\
  \bibnamefont {Naldesi}}, \ and\ \bibinfo {author} {\bibfnamefont {Tommaso}\
  \bibnamefont {Roscilde}},\ }\bibfield  {title} {\enquote {\bibinfo {title}
  {Multispeed prethermalization in quantum spin models with power-law decaying
  interactions},}\ }\href {\doibase 10.1103/PhysRevLett.120.050401} {\bibfield
  {journal} {\bibinfo  {journal} {Physical Review Letters}\ }\textbf {\bibinfo
  {volume} {120}},\ \bibinfo {pages} {050401} (\bibinfo {year}
  {2018})}\BibitemShut {NoStop}%
\bibitem [{\citenamefont {Vanderstraeten}\ \emph {et~al.}(2015)\citenamefont
  {Vanderstraeten}, \citenamefont {Verstraete},\ and\ \citenamefont
  {Haegeman}}]{Vanderstraeten2015a}%
  \BibitemOpen
  \bibfield  {author} {\bibinfo {author} {\bibfnamefont {L.}~\bibnamefont
  {Vanderstraeten}}, \bibinfo {author} {\bibfnamefont {F.}~\bibnamefont
  {Verstraete}}, \ and\ \bibinfo {author} {\bibfnamefont {J.}~\bibnamefont
  {Haegeman}},\ }\bibfield  {title} {\enquote {\bibinfo {title} {{Scattering
  particles in quantum spin chains}},}\ }\href {\doibase
  10.1103/PhysRevB.92.125136} {\bibfield  {journal} {\bibinfo  {journal}
  {Physical Review B}\ }\textbf {\bibinfo {volume} {92}},\ \bibinfo {pages}
  {125136} (\bibinfo {year} {2015})}\BibitemShut {NoStop}%
\bibitem [{\citenamefont {Vanderstraeten}\ \emph {et~al.}(2016)\citenamefont
  {Vanderstraeten}, \citenamefont {Haegeman}, \citenamefont {Verstraete},\ and\
  \citenamefont {Poilblanc}}]{Vanderstraeten2016}%
  \BibitemOpen
  \bibfield  {author} {\bibinfo {author} {\bibfnamefont {L.}~\bibnamefont
  {Vanderstraeten}}, \bibinfo {author} {\bibfnamefont {J.}~\bibnamefont
  {Haegeman}}, \bibinfo {author} {\bibfnamefont {F.}~\bibnamefont
  {Verstraete}}, \ and\ \bibinfo {author} {\bibfnamefont {D.}~\bibnamefont
  {Poilblanc}},\ }\bibfield  {title} {\enquote {\bibinfo {title}
  {{Quasiparticle interactions in frustrated Heisenberg chains}},}\ }\href
  {\doibase 10.1103/PhysRevB.93.235108} {\bibfield  {journal} {\bibinfo
  {journal} {Physical Review B}\ }\textbf {\bibinfo {volume} {93}},\ \bibinfo
  {pages} {235108} (\bibinfo {year} {2016})}\BibitemShut {NoStop}%
\bibitem [{\citenamefont {Verstraete}\ \emph {et~al.}(2008)\citenamefont
  {Verstraete}, \citenamefont {Murg},\ and\ \citenamefont
  {Cirac}}]{Verstraete2008a}%
  \BibitemOpen
  \bibfield  {author} {\bibinfo {author} {\bibfnamefont {F.}~\bibnamefont
  {Verstraete}}, \bibinfo {author} {\bibfnamefont {V.}~\bibnamefont {Murg}}, \
  and\ \bibinfo {author} {\bibfnamefont {J.~I.}\ \bibnamefont {Cirac}},\
  }\bibfield  {title} {\enquote {\bibinfo {title} {{Matrix product states,
  projected entangled pair states, and variational renormalization group
  methods for quantum spin systems}},}\ }\href {\doibase
  10.1080/14789940801912366} {\bibfield  {journal} {\bibinfo  {journal}
  {Advances in Physics}\ }\textbf {\bibinfo {volume} {57}},\ \bibinfo {pages}
  {143--224} (\bibinfo {year} {2008})}\BibitemShut {NoStop}%
\bibitem [{\citenamefont {Schollw{\"{o}}ck}(2011)}]{Schollwock2011a}%
  \BibitemOpen
  \bibfield  {author} {\bibinfo {author} {\bibfnamefont {U.}~\bibnamefont
  {Schollw{\"{o}}ck}},\ }\bibfield  {title} {\enquote {\bibinfo {title} {{The
  density-matrix renormalization group in the age of matrix product states}},}\
  }\href {\doibase 10.1016/j.aop.2010.09.012} {\bibfield  {journal} {\bibinfo
  {journal} {Annals of Physics}\ }\textbf {\bibinfo {volume} {326}},\ \bibinfo
  {pages} {96--192} (\bibinfo {year} {2011})}\BibitemShut {NoStop}%
\bibitem [{\citenamefont {Or{\'{u}}s}(2014)}]{Orus2013}%
  \BibitemOpen
  \bibfield  {author} {\bibinfo {author} {\bibfnamefont {R.}~\bibnamefont
  {Or{\'{u}}s}},\ }\bibfield  {title} {\enquote {\bibinfo {title} {{A practical
  introduction to tensor networks: Matrix product states and projected
  entangled pair states}},}\ }\href {\doibase 10.1016/j.aop.2014.06.013}
  {\bibfield  {journal} {\bibinfo  {journal} {Annals of Physics}\ }\textbf
  {\bibinfo {volume} {349}},\ \bibinfo {pages} {117--158} (\bibinfo {year}
  {2014})}\BibitemShut {NoStop}%
\bibitem [{\citenamefont {Zauner-Stauber}\ \emph {et~al.}(2017)\citenamefont
  {Zauner-Stauber}, \citenamefont {Vanderstraeten}, \citenamefont {Fishman},
  \citenamefont {Verstraete},\ and\ \citenamefont
  {Haegeman}}]{Zauner-Stauber2017}%
  \BibitemOpen
  \bibfield  {author} {\bibinfo {author} {\bibfnamefont {V.}~\bibnamefont
  {Zauner-Stauber}}, \bibinfo {author} {\bibfnamefont {L.}~\bibnamefont
  {Vanderstraeten}}, \bibinfo {author} {\bibfnamefont {M.~T.}\ \bibnamefont
  {Fishman}}, \bibinfo {author} {\bibfnamefont {F.}~\bibnamefont {Verstraete}},
  \ and\ \bibinfo {author} {\bibfnamefont {J.}~\bibnamefont {Haegeman}},\
  }\bibfield  {title} {\enquote {\bibinfo {title} {{Variational optimization
  algorithms for uniform matrix product states}},}\ }\href
  {http://arxiv.org/abs/1701.07035} {\  (\bibinfo {year} {2017})},\ \Eprint
  {http://arxiv.org/abs/1701.07035} {arXiv:1701.07035} \BibitemShut {NoStop}%
\bibitem [{\citenamefont {Haegeman}\ \emph {et~al.}(2012)\citenamefont
  {Haegeman}, \citenamefont {Pirvu}, \citenamefont {Weir}, \citenamefont
  {Cirac}, \citenamefont {Osborne}, \citenamefont {Verschelde},\ and\
  \citenamefont {Verstraete}}]{Haegeman2012a}%
  \BibitemOpen
  \bibfield  {author} {\bibinfo {author} {\bibfnamefont {J.}~\bibnamefont
  {Haegeman}}, \bibinfo {author} {\bibfnamefont {B.}~\bibnamefont {Pirvu}},
  \bibinfo {author} {\bibfnamefont {D.~J.}\ \bibnamefont {Weir}}, \bibinfo
  {author} {\bibfnamefont {J.~I.}\ \bibnamefont {Cirac}}, \bibinfo {author}
  {\bibfnamefont {T.~J.}\ \bibnamefont {Osborne}}, \bibinfo {author}
  {\bibfnamefont {H.}~\bibnamefont {Verschelde}}, \ and\ \bibinfo {author}
  {\bibfnamefont {F.}~\bibnamefont {Verstraete}},\ }\bibfield  {title}
  {\enquote {\bibinfo {title} {{Variational matrix product ansatz for
  dispersion relations}},}\ }\href {\doibase 10.1103/PhysRevB.85.100408}
  {\bibfield  {journal} {\bibinfo  {journal} {Physical Review B}\ }\textbf
  {\bibinfo {volume} {85}},\ \bibinfo {pages} {100408} (\bibinfo {year}
  {2012})}\BibitemShut {NoStop}%
\bibitem [{\citenamefont {Haldane}(1988)}]{Haldane1988}%
  \BibitemOpen
  \bibfield  {author} {\bibinfo {author} {\bibfnamefont {F.~D.~M.}\
  \bibnamefont {Haldane}},\ }\bibfield  {title} {\enquote {\bibinfo {title}
  {{Exact Jastrow-Gutzwiller resonating-valence-bond ground state of the
  spin-(1/2 antiferromagnetic Heisenberg chain with 1/r{\^{}}{\{}2{\}}
  exchange}},}\ }\href {\doibase 10.1103/PhysRevLett.60.635} {\bibfield
  {journal} {\bibinfo  {journal} {Physical Review Letters}\ }\textbf {\bibinfo
  {volume} {60}},\ \bibinfo {pages} {635--638} (\bibinfo {year}
  {1988})}\BibitemShut {NoStop}%
\bibitem [{\citenamefont {Shastry}(1988)}]{Shastry1988}%
  \BibitemOpen
  \bibfield  {author} {\bibinfo {author} {\bibfnamefont {B.~S.}\ \bibnamefont
  {Shastry}},\ }\bibfield  {title} {\enquote {\bibinfo {title} {{Exact solution
  of an S=1/2 Heisenberg antiferromagnetic chain with long-ranged
  interactions}},}\ }\href {\doibase 10.1103/PhysRevLett.60.639} {\bibfield
  {journal} {\bibinfo  {journal} {Physical Review Letters}\ }\textbf {\bibinfo
  {volume} {60}},\ \bibinfo {pages} {639--642} (\bibinfo {year}
  {1988})}\BibitemShut {NoStop}%
\bibitem [{\citenamefont {Faddeev}\ and\ \citenamefont
  {Takhtajan}(1981)}]{Faddeev1981}%
  \BibitemOpen
  \bibfield  {author} {\bibinfo {author} {\bibfnamefont {L.~D.}\ \bibnamefont
  {Faddeev}}\ and\ \bibinfo {author} {\bibfnamefont {L.~A.}\ \bibnamefont
  {Takhtajan}},\ }\bibfield  {title} {\enquote {\bibinfo {title} {{What is the
  spin of a spin wave?}}}\ }\href {\doibase 10.1016/0375-9601(81)90335-2}
  {\bibfield  {journal} {\bibinfo  {journal} {Physics Letters A}\ }\textbf
  {\bibinfo {volume} {85}},\ \bibinfo {pages} {375--377} (\bibinfo {year}
  {1981})}\BibitemShut {NoStop}%
\bibitem [{\citenamefont {Laflorencie}\ \emph {et~al.}(2005)\citenamefont
  {Laflorencie}, \citenamefont {Affleck},\ and\ \citenamefont
  {Berciu}}]{Laflorencie2005}%
  \BibitemOpen
  \bibfield  {author} {\bibinfo {author} {\bibfnamefont {N.}~\bibnamefont
  {Laflorencie}}, \bibinfo {author} {\bibfnamefont {I.}~\bibnamefont
  {Affleck}}, \ and\ \bibinfo {author} {\bibfnamefont {M.}~\bibnamefont
  {Berciu}},\ }\bibfield  {title} {\enquote {\bibinfo {title} {{Critical
  phenomena and quantum phase transition in long range Heisenberg
  antiferromagnetic chains}},}\ }\href {\doibase
  10.1088/1742-5468/2005/12/P12001} {\bibfield  {journal} {\bibinfo  {journal}
  {Journal of Statistical Mechanics: Theory and Experiment}\ }\textbf {\bibinfo
  {volume} {2005}},\ \bibinfo {pages} {P12001--P12001} (\bibinfo {year}
  {2005})}\BibitemShut {NoStop}%
\bibitem [{\citenamefont {Gong}\ \emph {et~al.}(2016)\citenamefont {Gong},
  \citenamefont {Maghrebi}, \citenamefont {Hu}, \citenamefont {Wall},
  \citenamefont {Foss-Feig},\ and\ \citenamefont {Gorshkov}}]{Gong2016}%
  \BibitemOpen
  \bibfield  {author} {\bibinfo {author} {\bibfnamefont {Z.-X.}\ \bibnamefont
  {Gong}}, \bibinfo {author} {\bibfnamefont {M.~F.}\ \bibnamefont {Maghrebi}},
  \bibinfo {author} {\bibfnamefont {A.}~\bibnamefont {Hu}}, \bibinfo {author}
  {\bibfnamefont {M.~L.}\ \bibnamefont {Wall}}, \bibinfo {author}
  {\bibfnamefont {M.}~\bibnamefont {Foss-Feig}}, \ and\ \bibinfo {author}
  {\bibfnamefont {A.~V.}\ \bibnamefont {Gorshkov}},\ }\bibfield  {title}
  {\enquote {\bibinfo {title} {Topological phases with long-range
  interactions},}\ }\href {\doibase 10.1103/PhysRevB.93.041102} {\bibfield
  {journal} {\bibinfo  {journal} {Physical Review B}\ }\textbf {\bibinfo
  {volume} {93}},\ \bibinfo {pages} {041102} (\bibinfo {year}
  {2016})}\BibitemShut {NoStop}%
\bibitem [{\citenamefont {Bernevig}\ \emph {et~al.}(2001)\citenamefont
  {Bernevig}, \citenamefont {Giuliano},\ and\ \citenamefont
  {Laughlin}}]{Bernevig2001}%
  \BibitemOpen
  \bibfield  {author} {\bibinfo {author} {\bibfnamefont {B.}~\bibnamefont
  {Bernevig}}, \bibinfo {author} {\bibfnamefont {D.}~\bibnamefont {Giuliano}},
  \ and\ \bibinfo {author} {\bibfnamefont {R.}~\bibnamefont {Laughlin}},\
  }\bibfield  {title} {\enquote {\bibinfo {title} {{Coordinate representation
  of the two-spinon wave function and spinon interaction in the Haldane-Shastry
  model}},}\ }\href {\doibase 10.1103/PhysRevB.64.024425} {\bibfield  {journal}
  {\bibinfo  {journal} {Physical Review B}\ }\textbf {\bibinfo {volume} {64}},\
  \bibinfo {pages} {024425} (\bibinfo {year} {2001})}\BibitemShut {NoStop}%
\bibitem [{\citenamefont {Sachdev}(2011)}]{Sachdev2011}%
  \BibitemOpen
  \bibfield  {author} {\bibinfo {author} {\bibfnamefont {S.}~\bibnamefont
  {Sachdev}},\ }\href@noop {} {\emph {\bibinfo {title} {{Quantum Phase
  Transitions}}}}\ (\bibinfo  {publisher} {Cambridge University Press},\
  \bibinfo {year} {2011})\BibitemShut {NoStop}%
\bibitem [{\citenamefont {Deng}\ \emph {et~al.}(2005)\citenamefont {Deng},
  \citenamefont {Porras},\ and\ \citenamefont {Cirac}}]{Deng2005}%
  \BibitemOpen
  \bibfield  {author} {\bibinfo {author} {\bibfnamefont {X.-L.}\ \bibnamefont
  {Deng}}, \bibinfo {author} {\bibfnamefont {D.}~\bibnamefont {Porras}}, \ and\
  \bibinfo {author} {\bibfnamefont {J.~I.}\ \bibnamefont {Cirac}},\ }\bibfield
  {title} {\enquote {\bibinfo {title} {Effective spin quantum phases in systems
  of trapped ions},}\ }\href {\doibase 10.1103/PhysRevA.72.063407} {\bibfield
  {journal} {\bibinfo  {journal} {Physical Review A}\ }\textbf {\bibinfo
  {volume} {72}},\ \bibinfo {pages} {063407} (\bibinfo {year}
  {2005})}\BibitemShut {NoStop}%
\bibitem [{\citenamefont {Vodola}\ \emph {et~al.}(2016)\citenamefont {Vodola},
  \citenamefont {Lepori}, \citenamefont {Ercolessi},\ and\ \citenamefont
  {Pupillo}}]{Vodola2016}%
  \BibitemOpen
  \bibfield  {author} {\bibinfo {author} {\bibfnamefont {D.}~\bibnamefont
  {Vodola}}, \bibinfo {author} {\bibfnamefont {L.}~\bibnamefont {Lepori}},
  \bibinfo {author} {\bibfnamefont {E.}~\bibnamefont {Ercolessi}}, \ and\
  \bibinfo {author} {\bibfnamefont {G.}~\bibnamefont {Pupillo}},\ }\bibfield
  {title} {\enquote {\bibinfo {title} {Long-range ising and kitaev models:
  phases, correlations and edge modes},}\ }\href
  {http://stacks.iop.org/1367-2630/18/i=1/a=015001} {\bibfield  {journal}
  {\bibinfo  {journal} {New Journal of Physics}\ }\textbf {\bibinfo {volume}
  {18}},\ \bibinfo {pages} {015001} (\bibinfo {year} {2016})}\BibitemShut
  {NoStop}%
\bibitem [{\citenamefont {Halimeh}\ and\ \citenamefont
  {Zauner-Stauber}(2017)}]{Halimeh2017}%
  \BibitemOpen
  \bibfield  {author} {\bibinfo {author} {\bibfnamefont {J.~C.}\ \bibnamefont
  {Halimeh}}\ and\ \bibinfo {author} {\bibfnamefont {V.}~\bibnamefont
  {Zauner-Stauber}},\ }\bibfield  {title} {\enquote {\bibinfo {title}
  {Dynamical phase diagram of quantum spin chains with long-range
  interactions},}\ }\href {\doibase 10.1103/PhysRevB.96.134427} {\bibfield
  {journal} {\bibinfo  {journal} {Physical Review B}\ }\textbf {\bibinfo
  {volume} {96}},\ \bibinfo {pages} {134427} (\bibinfo {year}
  {2017})}\BibitemShut {NoStop}%
\bibitem [{\citenamefont {Homrighausen}\ \emph {et~al.}(2017)\citenamefont
  {Homrighausen}, \citenamefont {Abeling}, \citenamefont {Zauner-Stauber},\
  and\ \citenamefont {Halimeh}}]{Homrighausen2017}%
  \BibitemOpen
  \bibfield  {author} {\bibinfo {author} {\bibfnamefont {I.}~\bibnamefont
  {Homrighausen}}, \bibinfo {author} {\bibfnamefont {N.~O.}\ \bibnamefont
  {Abeling}}, \bibinfo {author} {\bibfnamefont {V.}~\bibnamefont
  {Zauner-Stauber}}, \ and\ \bibinfo {author} {\bibfnamefont {J.~C.}\
  \bibnamefont {Halimeh}},\ }\bibfield  {title} {\enquote {\bibinfo {title}
  {Anomalous dynamical phase in quantum spin chains with long-range
  interactions},}\ }\href {\doibase 10.1103/PhysRevB.96.104436} {\bibfield
  {journal} {\bibinfo  {journal} {Physical Review B}\ }\textbf {\bibinfo
  {volume} {96}},\ \bibinfo {pages} {104436} (\bibinfo {year}
  {2017})}\BibitemShut {NoStop}%
\bibitem [{\citenamefont {Cardy}(1996)}]{Cardy1996}%
  \BibitemOpen
  \bibfield  {author} {\bibinfo {author} {\bibfnamefont {J.}~\bibnamefont
  {Cardy}},\ }\href@noop {} {\emph {\bibinfo {title} {{Scaling and
  Renormalization in Statistical Physics}}}}\ (\bibinfo  {publisher} {Cambridge
  University Press},\ \bibinfo {year} {1996})\BibitemShut {NoStop}%
\bibitem [{\citenamefont {Vanderstraeten}(2016)}]{Vanderstraeten2017}%
  \BibitemOpen
  \bibfield  {author} {\bibinfo {author} {\bibfnamefont {L.}~\bibnamefont
  {Vanderstraeten}},\ }\href
  {http://quantumtensor.pks.mpg.de/index.php/school/} {\enquote {\bibinfo
  {title} {Tangent-space methods for matrix product states},}\ } (\bibinfo
  {year} {2016})\BibitemShut {NoStop}%
\end{thebibliography}%

\allowdisplaybreaks[1]
\begin{center}
\textbf{\large SUPPLEMENTAL MATERIAL}
\end{center}

In this supplemental material, we provide (i) a calculation of the spectrum of the extended long-range Ising model, (ii) a mean-field argument for the irrelevance of the long-range interactions in the Heisenberg model, and (iii) the details for implementing the quasiparticle ansatz.

\section{Spectrum of the extended long-range Ising model}

The extended Ising model has the Hamiltonian
\begin{equation}
H = - \sum_{i<j} \frac{1}{(j-i)^\alpha} \sigma^z_i \left( \prod_{i<n<j} \sigma^x_n \right) \sigma^z_j - \lambda \sum_i \sigma^x_i,
\end{equation}
which can be mapped to a fermionic model using the Jordan-Wigner transformation
\begin{align}
\sigma^x_i &= 1-2 a^\dagger_i a_i \\
\sigma^z_i &=- \prod_{j<i}\left( 1-2 a^\dagger_j a_j \right) \left(a_i + a^\dagger_i \right).
\end{align}
After Fourier transforming, this yields
\begin{multline}
H=E_{cl} + \int \frac{dp}{2\pi} \\ \left( A(p) c^\dagger_p c_p + B(p) (c_{-p} c_p - c^\dagger_p c^\dagger_{-p}) \right),
\end{multline}
where
\begin{align}
& A(p)=2\lambda - 2\sum^\infty_{d=1} cos(dp)d^{-\alpha}
\\
& B(p)=-i\sum^\infty_{d=1}sin(dp)d^{-\alpha}.
\end{align}
This Hamiltonian is quadratic and can be solved using a Bogoliubov transformation $c_k=u_k b_k-iv_k b^\dagger_{-k}$ where $u_k^2+v_k^2=1$ , both $u_k$ and $v_k$ can assumed to be real valued and $u_k=u_{-k}$, $v_k=-v_{-k}$. The Hamiltonian simplifies to
\begin{equation}
H_{\text{ILC}} = E_0 + \int \frac{\d p}{2\pi} \omega(p) b_p\dag b_p
\end{equation}
with the dispersion relation
\begin{align}
& \omega(p) = 2\sqrt{(\lambda-C_\alpha(p))^2-S_\alpha(p)^2} ,\\
& C_\alpha(p) = \sum_{r=1}^L\frac{\cos(pr)}{r^{\alpha}}, \quad 
S_\alpha(p) = \sum_{r=1}^L\frac{\sin(kr)}{r^{\alpha}},
\end{align}

\section{Long-range Heisenberg model}

The relatively small influence of the long-range interaction in the Heisenberg model can be understood using an argument given by Cardy \cite{Cardy1996, Laflorencie2005}. We start by rewriting the Hamiltonian as a perturbation on the the nearest neighbour limit,
\begin{equation}
H = \sum_{i} \left( \sigma^x_i \sigma^x_{i+1}+\sigma^y_i \sigma^y_{i+1}+\sigma^z_i \sigma^z_{i+1} \right) + H_{\text{pert}}
\end{equation}
with
\begin{equation}
H_{\text{pert}}  = \sum_{i>j+1} \frac{\sigma^x_i \sigma^x_j+\sigma^y_i \sigma^y_j+\sigma^z_i \sigma^z_j}{(i-j)^{\alpha}}.
\end{equation}
The mean-field correction to the ground-state energy will be given by
\begin{equation}
\delta E = \sum_{i>j+1} \frac{ \braket{\sigma^x_i \sigma^x_j+\sigma^y_i \sigma^y_j+\sigma^z_i \sigma^z_j} }{(i-j)^{\alpha}}
\end{equation}
where the expectation value is evaluated in the unperturbed system and is known to scale as
\begin{equation}
\braket{\sigma^x_i \sigma^x_j+\sigma^y_i \sigma^y_j+\sigma^z_i \sigma^z_j} \sim \frac{(-1)^{i-j}}{|i-j|}.
\end{equation}
On a finite system of length $L$, the correction per site therefore scales as
\begin{equation}
\delta E \sim \int_1^{L/2} (2r)^{-\alpha -1}-(2r+1)^{-\alpha-1} dr \sim L^{-\alpha (\alpha + 1)}.
\end{equation}
We can compare this with finite-size corrections to the unperturbed system which are known to scale as $L^{-2}$ to lowest order. This tells us that the long-range perturbation is dominated by the finite size corrections for $\alpha > 1$ and is probably irrelevant.

%

\section{The quasiparticle ansatz: technical details}

In this section we write down the formulas needed for implementing the quasiparticle ansatz for a given long-range Hamiltonian
\begin{equation}
H = \sum_k c_k \sum_{i<j} \mu_k^{|i-j|} (P_1)_i (P_2)_j  + \sum_i Q_i.
\end{equation}
We use the vumps algorithm for optimizing the matrix product state approximation for the ground state, as explained in Ref.~\onlinecite{Zauner-Stauber2017}. This yields a ground state in the so-called mixed canonical form \cite{Vanderstraeten2017}
\begin{equation} 
\ket{\Psi(A)} =  \dots
\begin{diagram}
\draw (0.5,1.5) -- (1,1.5); 
\draw[rounded corners] (1,2) rectangle (2,1);
\draw (1.5,1.5) node (X) {$A_L$};
\draw (2,1.5) -- (3,1.5); 
\draw[rounded corners] (3,2) rectangle (4,1);
\draw (3.5,1.5) node {$A_L$};
\draw (4,1.5) -- (5,1.5);
\draw[rounded corners] (5,2) rectangle (6,1);
\draw (5.5,1.5) node {$A_C$};
\draw (6,1.5) -- (7,1.5); 
\draw[rounded corners] (7,2) rectangle (8,1);
\draw (7.5,1.5) node {$A_R$};
\draw (8,1.5) -- (9,1.5); 
\draw[rounded corners] (9,2) rectangle (10,1);
\draw (9.5,1.5) node {$A_R$};
\draw (10,1.5) -- (10.5,1.5);
\draw (1.5,1) -- (1.5,.5); \draw (3.5,1) -- (3.5,.5); \draw (5.5,1) -- (5.5,.5);
\draw (7.5,1) -- (7.5,.5); \draw (9.5,1) -- (9.5,.5); 
\end{diagram} \dots.
\end{equation}
We use the usual parametrization for the quasiparticle excitation \cite{Vanderstraeten2017}, so that, in order to variationally optimize the tensor $B$, we need to compute an overlap of the form
\begin{equation}
\bra{\Phi_{p'}(B')} H \ket{\Phi_p(B)} = 2\pi\delta(p-p') y.
\end{equation}
We will split up the expression for $y$ in a number of different contribution as follows.

\begin{widetext}

\par Let us first introduce the following tensors
\begin{equation}
\begin{diagram}
\draw (0,0) circle (.5);
\draw (0,0) node (X) {$O_L^k$};
\draw (0,0.5) edge[out=90,in=180] (1,1.5);
\draw (0,-0.5) edge[out=270,in=180] (1,-1.5);
\end{diagram} =
\begin{diagram}
\draw (1,-1.5) edge[out=180,in=180] (1,1.5);
\draw[rounded corners] (1,2) rectangle (2,1);
\draw[rounded corners] (1,-1) rectangle (2,-2);
\draw (1.5,0) circle (.5);
\draw (1.5,0) node {$P_1$};
\draw (1.5,1.5) node {$A_L$};
\draw (1.5,-1.5) node {$\bar{A}_L$};
\draw (1.5,1) -- (1.5,.5);
\draw (1.5,-1) -- (1.5,-.5);
\draw (2,1.5) -- (3,1.5); \draw (2,-1.5) -- (3,-1.5);
\draw[rounded corners] (3,2) rectangle (7,-2);
\draw (5,0) node {$\left(\one-\mu_k T_L\right)^{-1}$};
\draw (7,1.5) -- (7.5,1.5);  \draw (7,-1.5) -- (7.5,-1.5);
\end{diagram}
\; , \qquad
\begin{diagram}
\draw (2,1.5) edge[out=0,in=90] (3,0.5);
\draw (3,0) circle (.5);
\draw (3,-.5) edge[out = -90, in = 0] (2, -1.5);
\draw (3,0) node (X) {$O_R^k$}; 
\end{diagram} = 
\begin{diagram}
\draw (1.5,1.5) -- (2,1.5); \draw (1.5,-1.5) -- (2,-1.5);
\draw[rounded corners] (2,2) rectangle (6,-2);
\draw (4,0) node {$\left(\one-\mu_k T_R\right)^{-1}$};
\draw (6,1.5) -- (7,1.5);  \draw (6,-1.5) -- (7,-1.5);
\draw[rounded corners] (7,2) rectangle (8,1);
\draw[rounded corners] (7,-1) rectangle (8,-2);
\draw (7.5,1.5) node {$A_R$};
\draw (7.5,-1.5) node {$\bar{A}_R$};
\draw (7.5,0) node {$P_2$};
\draw (7.5,1) -- (7.5,.5);
\draw (7.5,-1) -- (7.5,-.5);
\draw (7.5,0) circle (.5);
\draw (8,+1.5) edge[out = 0, in =0] (8, -1.5);
\end{diagram}
\end{equation}
capturing the infinite sum of terms of $P$ operators acting to the left and to the right, resp. With these definitions, we can define the infinite series of the full Hamiltonian
\begin{equation}
\begin{diagram}
\draw (0,0) circle (.5);
\draw (0,0) node (X) {$H_L$};
\draw (0,0.5) edge[out=90,in=180] (1,1.5);
\draw (0,-0.5) edge[out=270,in=180] (1,-1.5);
\end{diagram} =
\left( \sum_k c_k \mu_k \; 
\begin{diagram}
\draw (0,0) circle (.5);
\draw (0,0) node (X) {$O_L^k$};
\draw (0,0.5) edge[out=90,in=180] (1,1.5);
\draw (0,-0.5) edge[out=270,in=180] (1,-1.5);
\draw[rounded corners] (1,2) rectangle (2,1);
\draw[rounded corners] (1,-1) rectangle (2,-2);
\draw (1.5,0) circle (.5);
\draw (1.5,0) node {$P_2$};
\draw (1.5,1.5) node {$A_L$};
\draw (1.5,-1.5) node {$\bar{A}_L$};
\draw (1.5,1) -- (1.5,.5);
\draw (1.5,-1) -- (1.5,-.5);
\draw (2,1.5) -- (2.5,1.5); \draw (2,-1.5) -- (2.5,-1.5);
\end{diagram} + 
\begin{diagram}
\draw (1,-1.5) edge[out=180,in=180] (1,1.5);
\draw[rounded corners] (1,2) rectangle (2,1);
\draw[rounded corners] (1,-1) rectangle (2,-2);
\draw (1.5,0) circle (.5);
\draw (1.5,0) node {$Q$};
\draw (1.5,1.5) node {$A_L$};
\draw (1.5,-1.5) node {$\bar{A}_L$};
\draw (1.5,1) -- (1.5,.5);
\draw (1.5,-1) -- (1.5,-.5);
\draw (2,1.5) -- (2.5,1.5); \draw (2,-1.5) -- (2.5,-1.5);
\end{diagram} \right) 
\begin{diagram}
\draw (2.5,1.5) -- (3,1.5); \draw (2.5,-1.5) -- (3,-1.5);
\draw[rounded corners] (3,2) rectangle (7,-2);
\draw (5,0) node {$\left(\one- T_L\right)^{-1}$};
\draw (7,1.5) -- (7.5,1.5);  \draw (7,-1.5) -- (7.5,-1.5);
\end{diagram}
\end{equation}
and
\begin{equation}
\begin{diagram}
\draw (2,1.5) edge[out=0,in=90] (3,0.5);
\draw (3,0) circle (.5);
\draw (3,-.5) edge[out = -90, in = 0] (2, -1.5);
\draw (3,0) node (X) {$H_R$}; 
\end{diagram} = 
\begin{diagram}
\draw (1.5,1.5) -- (2,1.5); \draw (1.5,-1.5) -- (2,-1.5);
\draw[rounded corners] (2,2) rectangle (6,-2);
\draw (4,0) node {$\left(\one- T_R\right)^{-1}$};
\draw (6,1.5) -- (6.5,1.5);  \draw (6,-1.5) -- (6.5,-1.5);
\end{diagram}
\left( \sum_k c_k \mu_k \;
\begin{diagram}
\draw (6.5,1.5) -- (7,1.5);  \draw (6.5,-1.5) -- (7,-1.5);
\draw[rounded corners] (7,2) rectangle (8,1);
\draw[rounded corners] (7,-1) rectangle (8,-2);
\draw (7.5,1.5) node {$A_R$};
\draw (7.5,-1.5) node {$\bar{A}_R$};
\draw (7.5,0) node {$P_1$};
\draw (7.5,1) -- (7.5,.5);
\draw (7.5,-1) -- (7.5,-.5);
\draw (7.5,0) circle (.5);
\draw (8,1.5) edge[out=0,in=90] (9,0.5);
\draw (9,0) circle (.5); \draw (9,0) node (X) {$O_R^k$}; 
\draw (9,-.5) edge[out = -90, in = 0] (8, -1.5);
\end{diagram}
+
\begin{diagram}
\draw (6.5,1.5) -- (7,1.5);  \draw (6.5,-1.5) -- (7,-1.5);
\draw[rounded corners] (7,2) rectangle (8,1);
\draw[rounded corners] (7,-1) rectangle (8,-2);
\draw (7.5,1.5) node {$A_R$};
\draw (7.5,-1.5) node {$\bar{A}_R$};
\draw (7.5,0) node {$Q$};
\draw (7.5,1) -- (7.5,.5);
\draw (7.5,-1) -- (7.5,-.5);
\draw (7.5,0) circle (.5);
\draw (8,+1.5) edge[out = 0, in =0] (8, -1.5);
\end{diagram}
\right).
\end{equation}
The first part of the effective Hamiltonian contains the ``diagonal terms'' where $B$ and $B'$ are located on the same site:
\begin{equation}
y_1 =  
\begin{diagram}
\draw (0,0) circle (.5);
\draw (0,0) node (X) {$H_L$};
\draw (0,0.5) edge[out=90,in=180] (1,1.5);
\draw (0,-0.5) edge[out=270,in=180] (1,-1.5);
\draw[rounded corners] (1,2) rectangle (2,1);
\draw[rounded corners] (1,-1) rectangle (2,-2);
\draw (1.5,1.5) node {$B$};
\draw (1.5,-1.5) node {$\bar{B'}$};
\draw (1.5,1) -- (1.5,-1);
\draw (2,+1.5) edge[out = 0, in =0] (2, -1.5);
\end{diagram}
+
\begin{diagram}
\draw (1,-1.5) edge[out=180,in=180] (1,1.5);
\draw[rounded corners] (1,2) rectangle (2,1);
\draw[rounded corners] (1,-1) rectangle (2,-2);
\draw (1.5,1.5) node {$B$};
\draw (1.5,-1.5) node {$\bar{B'}$};
\draw (1.5,1) -- (1.5,-1);
\draw (2,1.5) edge[out=0,in=90] (3,0.5);
\draw (3,0) circle (.5);
\draw (3,-.5) edge[out = -90, in = 0] (2, -1.5);
\draw (3,0) node (X) {$H_R$};
\end{diagram}
+
\begin{diagram}
\draw (1,-1.5) edge[out=180,in=180] (1,1.5);
\draw[rounded corners] (1,2) rectangle (2,1);
\draw[rounded corners] (1,-1) rectangle (2,-2);
\draw (1.5,0) circle (.5);
\draw (1.5,0) node {$Q$};
\draw (1.5,1.5) node {$B$};
\draw (1.5,-1.5) node {$\bar{B'}$};
\draw (1.5,1) -- (1.5,.5); \draw (1.5,-1) -- (1.5,-.5);
\draw (2,+1.5) edge[out = 0, in =0] (2,-1.5);
\end{diagram} \\
+
\sum_k c_k \mu_k \left( 
\begin{diagram}
\draw (0,0) circle (.5);
\draw (0,0) node (X) {$O_L^k$};
\draw (0,0.5) edge[out=90,in=180] (1,1.5);
\draw (0,-0.5) edge[out=270,in=180] (1,-1.5);
\draw[rounded corners] (1,2) rectangle (2,1);
\draw[rounded corners] (1,-1) rectangle (2,-2);
\draw (1.5,1.5) node {$B$};
\draw (1.5,-1.5) node {$\bar{B'}$};
\draw (1.5,0) circle (.5);
\draw (1.5,1) -- (1.5,.5); \draw (1.5,-1) -- (1.5,-.5);
\draw (1.5,0) node {$P_2$};
\draw (2,+1.5) edge[out = 0, in =0] (2,-1.5);
\end{diagram}
+
\mu_k\; \begin{diagram}
\draw (0,0) circle (.5);
\draw (0,0) node (X) {$O_L^k$};
\draw (0,0.5) edge[out=90,in=180] (1,1.5);
\draw (0,-0.5) edge[out=270,in=180] (1,-1.5);
\draw[rounded corners] (1,2) rectangle (2,1);
\draw[rounded corners] (1,-1) rectangle (2,-2);
\draw (1.5,1.5) node {$B$};
\draw (1.5,-1.5) node {$\bar{B'}$};
\draw (1.5,1) -- (1.5,-1);
\draw (2,1.5) edge[out=0,in=90] (3,0.5);
\draw (3,0) circle (.5);
\draw (3,-.5) edge[out = -90, in = 0] (2, -1.5);
\draw (3,0) node (X) {$O_R^k$};
\end{diagram}
+
\begin{diagram}
\draw (1,-1.5) edge[out=180,in=180] (1,1.5);
\draw[rounded corners] (1,2) rectangle (2,1);
\draw[rounded corners] (1,-1) rectangle (2,-2);
\draw (1.5,1.5) node {$B$};
\draw (1.5,-1.5) node {$\bar{B'}$};
\draw (1.5,0) circle (.5);
\draw (1.5,1) -- (1.5,.5); \draw (1.5,-1) -- (1.5,-.5);
\draw (1.5,0) node {$P_1$};
\draw (2,1.5) edge[out=0,in=90] (3,0.5);
\draw (3,0) circle (.5);
\draw (3,-.5) edge[out = -90, in = 0] (2, -1.5);
\draw (3,0) node (X) {$O_R^k$};
\end{diagram}
\right).
\end{equation}
The second part contains the local contributions (where both $B$ and $B'$ live between $P_1$ and $P_2$)
\begin{multline}
y_2 =  
\sum_k c_k \mu_k\e^{-ip} 
\left( 
\begin{diagram}
\draw (1,-1.5) edge[out=180,in=180] (1,1.5);
\draw[rounded corners] (1,2) rectangle (2,1);
\draw[rounded corners] (1,-1) rectangle (2,-2);
\draw (1.5,1.5) node {$B$};
\draw (1.5,-1.5) node {$\bar{A}_L$};
\draw (1.5,0) circle (.5);
\draw (1.5,0) node {$P_1$};
\draw (1.5,1) -- (1.5,.5); \draw (1.5,-.5) -- (1.5,-1);
\draw (2,1.5) -- (2.5,1.5); \draw (2,-1.5) -- (2.5,-1.5);
\end{diagram}
+
\mu_k \; \begin{diagram}
\draw (0,0) circle (.5);
\draw (0,0) node (X) {$O_L^k$};
\draw (0,0.5) edge[out=90,in=180] (1,1.5);
\draw (0,-0.5) edge[out=270,in=180] (1,-1.5);
\draw[rounded corners] (1,2) rectangle (2,1);
\draw[rounded corners] (1,-1) rectangle (2,-2);
\draw (1.5,1.5) node {$B$};
\draw (1.5,-1.5) node {$\bar{A}_L$};
\draw (1.5,1) -- (1.5,-1);
\draw (2,1.5) -- (2.5,1.5); \draw (2,-1.5) -- (2.5,-1.5);
\end{diagram}
\right) 
\begin{diagram}
\draw (1.5,1.5) -- (2,1.5);  \draw (1.5,-1.5) -- (2,-1.5);
\draw[rounded corners] (2,2) rectangle (8,-2);
\draw (5,0) node {$\left(\one- \mu_k \e^{-ip}T_L^R\right)^{-1}$};
\draw (8,1.5) -- (8.5,1.5);  \draw (8,-1.5) -- (8.5,-1.5);
\end{diagram}
\left( 
\mu_k \begin{diagram}
\draw (6.5,1.5) -- (7,1.5);  \draw (6.5,-1.5) -- (7,-1.5);
\draw[rounded corners] (7,2) rectangle (8,1);
\draw[rounded corners] (7,-1) rectangle (8,-2);
\draw (7.5,1.5) node {$A_R$};
\draw (7.5,-1.5) node {$\bar{B}'$};
\draw (7.5,1) -- (7.5,-1);
\draw (8,1.5) edge[out=0,in=90] (9,0.5);
\draw (9,0) circle (.5);
\draw (9,-.5) edge[out = -90, in = 0] (8, -1.5);
\draw (9,0) node (X) {$O_R^k$};
\end{diagram}
+
\begin{diagram}
\draw (6.5,1.5) -- (7,1.5);  \draw (6.5,-1.5) -- (7,-1.5);
\draw[rounded corners] (7,2) rectangle (8,1);
\draw[rounded corners] (7,-1) rectangle (8,-2);
\draw (7.5,1.5) node {$A_R$};
\draw (7.5,-1.5) node {$\bar{B}'$};
\draw (7.5,1) -- (7.5,.5); \draw (7.5,-.5) -- (7.5,-1);
\draw (7.5,0) circle (.5);
\draw (7.5,0) node {$P_2$};
\draw (8,+1.5) edge[out = 0, in =0] (8,-1.5);
\end{diagram}
\right) \\
 + \sum_k c_k \mu_k\e^{+ip} 
\left( 
\begin{diagram}
\draw (1,-1.5) edge[out=180,in=180] (1,1.5);
\draw[rounded corners] (1,2) rectangle (2,1);
\draw[rounded corners] (1,-1) rectangle (2,-2);
\draw (1.5,1.5) node {$A_L$};
\draw (1.5,-1.5) node {$\bar{B'}$};
\draw (1.5,0) circle (.5);
\draw (1.5,0) node {$P_1$};
\draw (1.5,1) -- (1.5,.5); \draw (1.5,-.5) -- (1.5,-1);
\draw (2,1.5) -- (2.5,1.5); \draw (2,-1.5) -- (2.5,-1.5);
\end{diagram}
+
\mu_k \; \begin{diagram}
\draw (0,0) circle (.5);
\draw (0,0) node (X) {$O_L^k$};
\draw (0,0.5) edge[out=90,in=180] (1,1.5);
\draw (0,-0.5) edge[out=270,in=180] (1,-1.5);
\draw[rounded corners] (1,2) rectangle (2,1);
\draw[rounded corners] (1,-1) rectangle (2,-2);
\draw (1.5,1.5) node {$A_L$};
\draw (1.5,-1.5) node {$\bar{B}'$};
\draw (1.5,1) -- (1.5,-1);
\draw (2,1.5) -- (2.5,1.5); \draw (2,-1.5) -- (2.5,-1.5);
\end{diagram}
\right) 
\begin{diagram}
\draw (1.5,1.5) -- (2,1.5);  \draw (1.5,-1.5) -- (2,-1.5);
\draw[rounded corners] (2,2) rectangle (8,-2);
\draw (5,0) node {$\left(\one- \mu_k \e^{+ip}T_R^L\right)^{-1}$};
\draw (8,1.5) -- (8.5,1.5);  \draw (8,-1.5) -- (8.5,-1.5);
\end{diagram}
\left( 
\mu_k \begin{diagram}
\draw (6.5,1.5) -- (7,1.5);  \draw (6.5,-1.5) -- (7,-1.5);
\draw[rounded corners] (7,2) rectangle (8,1);
\draw[rounded corners] (7,-1) rectangle (8,-2);
\draw (7.5,1.5) node {$B$};
\draw (7.5,-1.5) node {$\bar{A}_R$};
\draw (7.5,1) -- (7.5,-1);
\draw (8,1.5) edge[out=0,in=90] (9,0.5);
\draw (9,0) circle (.5);
\draw (9,-.5) edge[out = -90, in = 0] (8, -1.5);
\draw (9,0) node (X) {$O_R^k$};
\end{diagram}
+
\begin{diagram}
\draw (6.5,1.5) -- (7,1.5);  \draw (6.5,-1.5) -- (7,-1.5);
\draw[rounded corners] (7,2) rectangle (8,1);
\draw[rounded corners] (7,-1) rectangle (8,-2);
\draw (7.5,1.5) node {$B$};
\draw (7.5,-1.5) node {$\bar{A}_R$};
\draw (7.5,1) -- (7.5,.5); \draw (7.5,-.5) -- (7.5,-1);
\draw (7.5,0) circle (.5);
\draw (7.5,0) node {$P_2$};
\draw (8,+1.5) edge[out = 0, in =0] (8,-1.5);
\end{diagram}
\right).
\end{multline}
The next part contains the contributions where $B$ is disconnected and travels to the right. We need the following matrix,
\begin{equation}
\begin{diagram}
\draw (2,1.5) edge[out=0,in=90] (3,0.5);
\draw (3,0) circle (.5);
\draw (3,-.5) edge[out = -90, in = 0] (2, -1.5);
\draw (3,0) node (X) {$R_B$}; 
\end{diagram} = 
\begin{diagram}
\draw (1.5,1.5) -- (2,1.5); \draw (1.5,-1.5) -- (2,-1.5);
\draw[rounded corners] (2,2) rectangle (6,-2);
\draw (4,0) node {$\left(\one-\e^{ip}T^L_R\right)^{-1}$};
\draw (6,1.5) -- (7,1.5);  \draw (6,-1.5) -- (7,-1.5);
\draw[rounded corners] (7,2) rectangle (8,1);
\draw[rounded corners] (7,-1) rectangle (8,-2);
\draw (7.5,1.5) node {$B$};
\draw (7.5,-1.5) node {$\bar{A}_R$};
\draw (7.5,1) -- (7.5,-1);
\draw (8,+1.5) edge[out = 0, in =0] (8, -1.5);
\end{diagram},
\end{equation}
and find the matrix elements as
\begin{multline}
y_3 =
\e^{ip} \; \begin{diagram}
\draw (0,0) circle (.5);
\draw (0,0) node (X) {$H_L$};
\draw (0,0.5) edge[out=90,in=180] (1,1.5);
\draw (0,-0.5) edge[out=270,in=180] (1,-1.5);
\draw[rounded corners] (1,2) rectangle (2,1);
\draw[rounded corners] (1,-1) rectangle (2,-2);
\draw (1.5,1.5) node {$A_L$};
\draw (1.5,-1.5) node {$\bar{B'}$};
\draw (1.5,1) -- (1.5,-1); 
\draw (2,1.5) edge[out=0,in=90] (3,0.5);
\draw (3,0) circle (.5);
\draw (3,-.5) edge[out = -90, in = 0] (2, -1.5);
\draw (3,0) node (X) {$R_B$};
\end{diagram}
+
\e^{ip} \begin{diagram}
\draw (1,-1.5) edge[out=180,in=180] (1,1.5);
\draw[rounded corners] (1,2) rectangle (2,1);
\draw[rounded corners] (1,-1) rectangle (2,-2);
\draw (1.5,0) circle (.5);
\draw (1.5,0) node {$Q$};
\draw (1.5,1.5) node {$A_L$};
\draw (1.5,-1.5) node {$\bar{B'}$};
\draw (1.5,1) -- (1.5,.5); \draw (1.5,-1) -- (1.5,-.5);
\draw (2,1.5) edge[out=0,in=90] (3,0.5);
\draw (3,0) circle (.5);
\draw (3,-.5) edge[out = -90, in = 0] (2, -1.5);
\draw (3,0) node (X) {$R_B$};
\end{diagram} 
+ \e^{ip} \sum_k c_k \mu_k  \;
\begin{diagram}
\draw (0,0) circle (.5);
\draw (0,0) node (X) {$O_L^k$};
\draw (0,0.5) edge[out=90,in=180] (1,1.5);
\draw (0,-0.5) edge[out=270,in=180] (1,-1.5);
\draw[rounded corners] (1,2) rectangle (2,1);
\draw[rounded corners] (1,-1) rectangle (2,-2);
\draw (1.5,1.5) node {$A_L$};
\draw (1.5,-1.5) node {$\bar{B'}$};
\draw (1.5,1) -- (1.5,.5); \draw (1.5,-1) -- (1.5,-.5);
\draw (1.5,0) circle (.5);
\draw (1.5,0) node {$P_2$};
\draw (2,1.5) edge[out=0,in=90] (3,0.5);
\draw (3,0) circle (.5);
\draw (3,-.5) edge[out = -90, in = 0] (2, -1.5);
\draw (3,0) node (X) {$R_B$};
\end{diagram} \\
+ \e^{2ip} \sum_k c_k \mu_k \left(
\mu_k \; \begin{diagram}
\draw (0,0) circle (.5);
\draw (0,0) node (X) {$O_L^k$};
\draw (0,0.5) edge[out=90,in=180] (1,1.5);
\draw (0,-0.5) edge[out=270,in=180] (1,-1.5);
\draw[rounded corners] (1,2) rectangle (2,1);
\draw[rounded corners] (1,-1) rectangle (2,-2);
\draw (1.5,1.5) node {$A_L$};
\draw (1.5,-1.5) node {$\bar{B'}$};
\draw (1.5,1) -- (1.5,-1);
\draw (2,1.5) -- (2.5,1.5);  \draw (2,-1.5) -- (2.5,-1.5);
\end{diagram}
+ \begin{diagram}
\draw (1,-1.5) edge[out=180,in=180] (1,1.5);
\draw[rounded corners] (1,2) rectangle (2,1);
\draw[rounded corners] (1,-1) rectangle (2,-2);
\draw (1.5,1.5) node {$A_L$};
\draw (1.5,-1.5) node {$\bar{B'}$};
\draw (1.5,1) -- (1.5,.5); \draw (1.5,-.5) -- (1.5,-1);
\draw (1.5,0) circle (.5);
\draw (1.5,0) node {$P_1$};
\draw (2,1.5) -- (2.5,1.5);  \draw (2,-1.5) -- (2.5,-1.5);
\end{diagram}
\right)
\begin{diagram}
\draw (0.5,1.5) -- (1,1.5);  \draw (0.5,-1.5) -- (1,-1.5);
\draw[rounded corners] (1,2) rectangle (7,-2);
\draw (4,0) node {$\left(\one- \mu_k \e^{+ip}T_R^L\right)^{-1}$};
\draw (7,1.5) -- (8,1.5);  \draw (7,-1.5) -- (8,-1.5);
\draw[rounded corners] (8,2) rectangle (9,1);
\draw[rounded corners] (8,-1) rectangle (9,-2);
\draw (8.5,1.5) node {$A_L$};
\draw (8.5,-1.5) node {$\bar{A}_R$};
\draw (8.5,1) -- (8.5,.5); \draw (8.5,-1) -- (8.5,-.5);
\draw (8.5,0) circle (.5);
\draw (8.5,0) node {$P_2$};
\draw (9,1.5) edge[out=0,in=90] (10,0.5);
\draw (10,0) circle (.5);
\draw (10,-.5) edge[out = -90, in = 0] (9, -1.5);
\draw (10,0) node (X) {$R_B$};
\end{diagram}\;.
\end{multline}
The last part contains the contributions where $B'$ is disconnected and travels to the right, whereas $B$ is between $P_1$ and $P_2$. Therefore, we define the following matrix
\begin{multline}
\begin{diagram}
\draw (0,0) circle (.5);
\draw (0,0) node (X) {$L$};
\draw (0,0.5) edge[out=90,in=180] (1,1.5);
\draw (0,-0.5) edge[out=270,in=180] (1,-1.5);
\end{diagram} =
\e^{-ip} \begin{diagram}
\draw (0,0) circle (.5);
\draw (0,0) node (X) {$H_L$};
\draw (0,0.5) edge[out=90,in=180] (1,1.5);
\draw (0,-0.5) edge[out=270,in=180] (1,-1.5);
\draw[rounded corners] (1,2) rectangle (2,1);
\draw[rounded corners] (1,-1) rectangle (2,-2);
\draw (1.5,1.5) node {$B$};
\draw (1.5,-1.5) node {$\bar{A}_L$};
\draw (1.5,1) -- (1.5,-1);
\draw (2,1.5) -- (2.5,1.5); \draw (2,-1.5) -- (2.5,-1.5);
\end{diagram}
+ \e^{-ip}
\begin{diagram}
\draw (1,-1.5) edge[out=180,in=180] (1,1.5);
\draw[rounded corners] (1,2) rectangle (2,1);
\draw[rounded corners] (1,-1) rectangle (2,-2);
\draw (1.5,0) circle (.5);
\draw (1.5,0) node {$Q$};
\draw (1.5,1.5) node {$B$};
\draw (1.5,-1.5) node {$\bar{A}_L$};
\draw (1.5,1) -- (1.5,.5);
\draw (1.5,-1) -- (1.5,-.5);
\draw (2,1.5) -- (2.5,1.5); \draw (2,-1.5) -- (2.5,-1.5);
\end{diagram} 
+ \e^{-ip} \sum_k c_k \mu_k \;
\begin{diagram}
\draw (0,0) circle (.5);
\draw (0,0) node (X) {$O_L^k$};
\draw (0,0.5) edge[out=90,in=180] (1,1.5);
\draw (0,-0.5) edge[out=270,in=180] (1,-1.5);
\draw[rounded corners] (1,2) rectangle (2,1);
\draw[rounded corners] (1,-1) rectangle (2,-2);
\draw (1.5,0) circle (.5);
\draw (1.5,0) node {$P_2$};
\draw (1.5,1.5) node {$B$};
\draw (1.5,-1.5) node {$\bar{A}_L$};
\draw (1.5,1) -- (1.5,.5);
\draw (1.5,-1) -- (1.5,-.5);
\draw (2,1.5) -- (2.5,1.5); \draw (2,-1.5) -- (2.5,-1.5);
\end{diagram} \\
 + \e^{-2ip} \sum_k c_k \mu_k  \left( \mu_k\;
\begin{diagram}
\draw (0,0) circle (.5);
\draw (0,0) node (X) {$O_L^k$};
\draw (0,0.5) edge[out=90,in=180] (1,1.5);
\draw (0,-0.5) edge[out=270,in=180] (1,-1.5);
\draw[rounded corners] (1,2) rectangle (2,1);
\draw[rounded corners] (1,-1) rectangle (2,-2);
\draw (1.5,1.5) node {$B$};
\draw (1.5,-1.5) node {$\bar{A}_L$};
\draw (1.5,1) -- (1.5,-1);
\draw (2,1.5) -- (2.5,1.5); \draw (2,-1.5) -- (2.5,-1.5);
\end{diagram}
+ 
\begin{diagram}
\draw (1,-1.5) edge[out=180,in=180] (1,1.5);
\draw[rounded corners] (1,2) rectangle (2,1);
\draw[rounded corners] (1,-1) rectangle (2,-2);
\draw (1.5,1.5) node {$B$};
\draw (1.5,-1.5) node {$\bar{A}_L$};
\draw (1.5,0) circle (.5);
\draw (1.5,0) node {$P_1$};
\draw (1.5,1) -- (1.5,.5); \draw (1.5,-.5) -- (1.5,-1);
\draw (2,1.5) -- (2.5,1.5); \draw (2,-1.5) -- (2.5,-1.5);
\end{diagram}
\right)
\begin{diagram}
\draw (0.5,1.5) -- (1,1.5);  \draw (0.5,-1.5) -- (1,-1.5);
\draw[rounded corners] (1,2) rectangle (7,-2);
\draw (4,0) node {$\left(\one- \mu_k \e^{-ip}T_L^R\right)^{-1}$};
\draw (7,1.5) -- (8,1.5);  \draw (7,-1.5) -- (8,-1.5);
\draw[rounded corners] (8,2) rectangle (9,1);
\draw[rounded corners] (8,-1) rectangle (9,-2);
\draw (8.5,0) circle (.5);
\draw (8.5,0) node {$P_2$};
\draw (8.5,1.5) node {$A_R$};
\draw (8.5,-1.5) node {$\bar{A}_L$};
\draw (8.5,1) -- (8.5,.5); \draw (8.5,-1) -- (8.5,-.5);
\draw (9,1.5) -- (9.5,1.5);  \draw (9,-1.5) -- (9.5,-1.5);
\end{diagram}\;.
\end{multline}
The last part of the effective Hamiltonian is then
\begin{equation}
y_4 = \begin{diagram}
\draw (-1,0) circle (.5);
\draw (-1,0) node (X) {$L$};
\draw (-1,0.5) edge[out=90,in=180] (0,1.5);
\draw (-1,-0.5) edge[out=270,in=180] (0,-1.5);
\draw[rounded corners] (0,2) rectangle (5,-2);
\draw (2.5,0) node {$\left(\one- \e^{-ip}T^R_L\right)^{-1}$};
\draw (5,1.5) -- (6,1.5);  \draw (5,-1.5) -- (6,-1.5);
\draw[rounded corners] (6,2) rectangle (7,1);
\draw[rounded corners] (6,-1) rectangle (7,-2);
\draw (6.5,1.5) node {$A_R$};
\draw (6.5,-1.5) node {$\bar{B'}$};
\draw (6.5,1) -- (6.5,-1);
\draw (7,+1.5) edge[out = 0, in =0] (7, -1.5);
\end{diagram}\;.
\end{equation}
The full Hamiltonian matrix is obtained by summing all four contributions,
\begin{equation}
y = y_1 + y_2 + y_3 + y_4.
\end{equation}

\end{widetext}

\phantom{Fake section}

\end{document}